\documentclass[]{emulateapj}
\usepackage{subfigure}
\usepackage{url}
\usepackage{hyperref}

\newcommand{\noprint}[1]{}

\newcommand{\ik}{{\it Kepler~}}

\newcommand{\vsini}{{$V \sin i$~}}
\newcommand{\teff}{$T_{\rm eff}$~}
\newcommand{\kms}{{km~$\rm s^{-1}$}}
\newcommand{\rstar}{{$R_\star$}~}
\newcommand{\mstar}{{$M_\star$}}
\newcommand{\logg}{{log(g)}~}
\newcommand{\mh}{{[M/H]}~}
\newcommand{\feh}{{[Fe/H]}~}

\newcommand{\spectra}{$H$- and $K$-band spectra of the Cool KOIs identified as dwarfs in this study, shifted to remove radial velocity.  The KOIs are ordered in increasing $T_{\rm eff}$.  The spectra are plotted on a log scale and multiplied by an arbitrary constant for the sake of ordering.  The spectra are available for download online using the Data Behind the Figure feature.}

\begin{document}

\title{Characterizing the Cool KOI\lowercase{s}. VI. $H$- and $K$-Band Spectra of {\it Kepler} M Dwarf Planet-Candidate Hosts}

\author{Philip S. Muirhead,\altaffilmark{1,11} Juliette Becker,\altaffilmark{2} Gregory A. Feiden,\altaffilmark{3} B{\'a}rbara~Rojas-Ayala,\altaffilmark{4} Andrew Vanderburg,\altaffilmark{5} Ellen M. Price,\altaffilmark{2} Rachel Thorp,\altaffilmark{2} Nicholas M. Law,\altaffilmark{6} Reed Riddle,\altaffilmark{2} Christoph Baranec,\altaffilmark{7} Katherine Hamren,\altaffilmark{8} Everett Schlawin,\altaffilmark{9} Kevin R. Covey,\altaffilmark{10} John Asher Johnson,\altaffilmark{5} James P. Lloyd\altaffilmark{9}}

\altaffiltext{1}{Department of Astronomy, Boston University, 725 Commonwealth Ave., Boston, MA  02215 USA; {\tt philipm@bu.edu}}
\altaffiltext{2}{California Institute of Technology, 1200 E. California Blvd., Pasadena, CA  91125 USA}
\altaffiltext{3}{Department of Physics and Astronomy, Uppsala University, Box 516, Uppsala 751 20, Sweden}
\altaffiltext{4}{Centro de Astrof'sica da Universidade do Porto, Rua das Estrelas, 4150-762 Oporto, Portugal}
\altaffiltext{5}{Harvard College Observatory, 60 Garden St., Cambridge, MA 02138 USA}
\altaffiltext{6}{Department of Physics and Astronomy, University of North Carolina at Chapel Hill, Chapel Hill, NC 27599-3255, USA}
\altaffiltext{7}{Institute for Astronomy, University of Hawai$\textquoteleft$i at M\={a}noa, Hilo, HI 96720-2700, USA}
\altaffiltext{8}{Department of Astronomy and Astrophysics, University of California, Santa Cruz CA 95064 USA}
\altaffiltext{9}{Department of Astronomy, Cornell University, Ithaca, NY  14583 USA}
\altaffiltext{10}{Lowell Observatory, 1400 W Mars Hill Rd., Flagstaff, AZ  86001 USA}
\altaffiltext{11}{Hubble Fellow}


\begin{abstract}

We present $H$- and $K$-band spectra for late-type {\it Kepler} Objects of Interest (the ``Cool KOIs''): low-mass stars with transiting-planet candidates discovered by NASA's {\it Kepler} Mission that are listed on the NASA Exoplanet Archive.  We acquired spectra of 103 Cool KOIs and used the indices and calibrations of Rojas-Ayala et al.\ to determine their spectral types, stellar effective temperatures and metallicities, significantly augmenting previously published values.  We interpolate our measured effective temperatures and metallicities onto evolutionary isochrones to determine stellar masses, radii, luminosities and distances, assuming the stars have settled onto the main-sequence.  As a choice of isochrones, we use a new suite of Dartmouth predictions that reliably include mid-to-late M dwarf stars.  We identify five M4V stars: KOI-961 (confirmed as {\it Kepler} 42), KOI-2704, KOI-2842, KOI-4290, and the secondary component to visual binary KOI-1725, which we call KOI-1725 B.  We also identify a peculiar star, KOI-3497, which has a Na and Ca lines consistent with a dwarf star but CO lines consistent with a giant.  Visible-wavelength adaptive optics imaging reveals two objects within a 1 arc second diameter; however, the objects' colors are peculiar.  The spectra and properties presented in this paper serve as a resource for prioritizing follow-up observations and planet validation efforts for the Cool KOIs, and are all available for download online using the ``data behind the figure'' feature.

\end{abstract}


\maketitle

\section{Introduction}

NASA's {\it Kepler} Mission has led exoplanet science into a new era \citep[]{Borucki2010b, Koch2010}.  By monitoring over 190,000 stars for transiting planets with unprecedented photometric precision over roughly 4 years, {\it Kepler} has discovered thousands of planet-candidates to date \citep[]{Borucki2011a, Borucki2011b, Batalha2013}, with over 100 now confirmed as genuine exoplanets \citep[e.g.][]{Ballard2011, Carter2012, Fressin2012, Johnson2012, Swift2013, Barclay2013}.  Though \ik can no longer achieving the same photometric precision due to problems with the spacecraft, more and more transiting planets will likely be discovered and confirmed in the archival data over the coming years.

Stars with planet-like transit signals detected by the {\it Kepler} analysis pipeline receive the designation {\it Kepler} Objects of Interest (KOIs).  With further study, the KOIs are given a disposition of either a ``candidate'' or ``false-positive'' \citep[][]{Jenkins2010, Christiansen2013}.  When confirmed as {\it bona fide} planetary transits, the KOIs are dispositioned as ``confirmed.''  Although most of the candidate KOIs have yet to be confirmed, {\it a priori} calculations indicate that over 90\% of all systems and 100\% of the multi-transiting systems are in fact {\it bona fide} exoplanet systems \citep{Morton2011, Lissauer2011, Fressin2013}.  

The sheer number of KOIs discovered by {\it Kepler} enables the calculation of planet-occurrence statistics versus stellar and planet properties \citep{Youdin2011, Howard2012, Mann2012, Dressing2013, Morton2013, Mann2013b, Mann2013c, Petigura2013, Marcy2014}.  One of the more intriguing results from these studies is the dependence of planet-occurrence on stellar mass.  \citet{Howard2012} found that for planets with orbital periods of less than 50 days and sizes of 2 to 4 Earth radii, planet-occurrence rises significantly with decreasing host star effective temperature.  This is in stark contrast to the frequency of short-period jovian-mass planets, which show the opposite trend consistent with core-accretion simulations \citep{Johnson2010, Adams2005}.  The results indicate that low-mass planets are very common around M dwarf stars, and recent statistical studies by \citet{Dressing2013} and \citet{Morton2013} appear to confirm that 1-Earth-radius planets orbit M dwarfs at a frequency of roughly 1 per star.

However, the measured transit depth in a given \ik light curve is proportional to the square of the planet-to-star radius ratio; therefore, the reported physical properties of transiting planet-candidates are only as accurate as the physical properties ascribed to their host stars.  For M dwarfs, masses and radii are typically determined by combining empirical mass-luminosity relationships \citep[e.g.][]{Delfosse2000} with empirical or model mass-radius relationships \citep[e.g.][]{Baraffe1998,Torres2010,Feiden2011,Feiden2012}.  Being relatively faint, {\it Kepler} target stars do not typically have {\it Hipparcos} or other archival astrometric parallax measurements, so luminosities and absolute photometric magnitudes are unknown.  In the absence of astrometric parallaxes, the burden of measuring physical properties of single M dwarfs typically falls on photometric colors and spectroscopy.

In their respective papers, \citet{Batalha2010} and \citet{Brown2011} describe the {\it Kepler} Input Catalog (KIC): a catalog of photometric measurements and stellar properties for targets in {\it Kepler}'s field of view.  The KIC contains Sloan-like $g$, $r$, $i$ and $z$ band photometry, as well as $J$-, $H$- and $K$-band photometry from 2MASS \citep{Cutri2003, Skrutskie2006}.  Using these measurements, \citet{Brown2011} determined stellar mass, radius, luminosity and metallicity using predictions from the atmospheric models of \citet{Castelli2004} and evolutionary models of \citet{Girardi2000}.  

For M dwarf stars, these atmospheric and evolutionary model predictions differ significantly from empirical observations \citep{Boyajian2012}.  Concerning the former, this can be attributed to complications arising from the treatment of the vast array of molecules at low temperatures. In the latter, the differences arise due to the inaccurate treatment of surface boundary conditions and the equation of state of the stellar interior. The \citet{Girardi2000} models assume a gray opacity for the stellar surface, an assumption that becomes increasingly less reliable with lower surface temperatures (below roughly 5000 $K$), where molecular absorption begins to dominate stellar opacities and convection in the outer layers becomes sufficiently non-adiabatic. Additionally, stellar interiors become partially degenerate toward lower masses, meaning the equation of state must account for non-ideal effects. This is especially pertinent for fully convective stars ($M_\star \lesssim 0.35 M_\Sun$). The same inaccuracies also apply to the Yonsei-Yale isochrones used for \ik planet hosts in \citet{Batalha2013}.

\citet{Brown2011} therefore state that the stellar properties for stars with effective temperatures of less than 3750 $K$ are ``untrustworthy.''  Indeed, in a previous letter \citet{Muirhead2012b} found large discrepancies between the stellar mass, radius and luminosity determinations of the KIC and determinations using $K$-band spectroscopy and the Dartmouth Stellar Evolutionary Database.  Similarly, \citet{Mann2012} found that the KIC effective temperature estimates for M dwarfs were 100 to 200 K higher than when estimated using optical spectroscopy, not to mention that many of the \ik targets classified as M dwarfs were in fact giant stars.  A recent paper by \citet{Pinsonneault2012} reported revised effective temperatures for the Kepler Targets using calibrations from the Sloan Digital Sky Survey and the Infrared Flux Method; however, they do not report revised temperatures for stars below 4000 K, which includes nearly all of the M dwarfs.

Recently, \citet{Dressing2013} reported new masses and radii for all the \ik M dwarfs observed as part of \ik's primary planet search program using a custom minimization of differences between the KIC photometry and photometric predictions from the the Dartmouth Stellar Evolutionary Database \citep{Dotter2008}.  They demonstrated their technique's accuracy on a set of nearby stars with parallaxes, and showed generally good agreement between their photometric mass determinations and the masses as determined by mass-luminosity relations.  However, when using photometry alone to determine stellar properties, effective temperature is degenerate with metallicity (either can make a star appear redder or bluer in a given color), and both temperature and metallicity play significant roles in determining M dwarf mass and radius on a given set of isochrones.  In fact, initially allowing metallicity to float freely in the fitting procedure, \citet{Dressing2013} found that their metallicity determinations were unrealistic.  To correct for this, they enforced a strong prior on the metallicity determinations for a given set of photometric measurements, confining them to a spread of metallicities representing the solar neighborhood.

In another study,  \citet{Mann2013b} reported metallicities and radii for the late K and M-type KOIs, and a control sample of Kepler targets that did not host transiting planet candidates, at the time of their study.  They found that the Cool KOIs that host Jupiter-sized transiting planet candidates were preferentially metal-rich, but that KOIs that host Neptune-sized and smaller planets and the Kepler target sample as a whole followed a metallicity distribution similar to the solar neighborhood.  This is in sharp contrast to the results of \citet{Schlaufman2011}, who found that all candidate host stars were preferentially metal rich, regardless of the planet candidate radii.  When determining the radii of the planet-hosting stars, \citet{Mann2013b} did not incorporate stellar metallicity.  Instead, they used an empirically-determined radius-to-temperature relationship from \citet{Boyajian2012}, who found little metallicity dependence to the relationship.  However, the majority of evolutionary isochrones, including the Dartmouth isochrones, predict significantly different radius-to-temperature relationships for stars of different metallicities.  The discrepancy between the \citet{Boyajian2012} results and isochrone predictions remains an important controversy to resolve.  We note, however, that the metallicity spread of the M dwarfs in the \citet{Boyajian2012} sample is small, due to being limited to the solar-neighborhood.  It is quite possible that the metallicity dependence to radius-to-temperature relationship only becomes apparent for significant metallicity differences.  With future accurate radius measurements for subdwarfs or so-called ``extreme'' and ``ultra'' subdwarfs \citep{Lepine2007}, we can better test the predicted metallicity dependence in the radius-to-temperature relationships.

In yet another study, \citet{Mann2013c} report effective temperatures for the late K and M-type KOIs using spectroscopic indices.  In this latest study, \citet{Mann2013c} calibrated spectroscopic indices directly to effective temperatures and stellar radii using the measurements from \citet{Boyajian2012}.  The difference between \citet{Mann2013b} and \citet{Mann2013c} is the inclusion of more KOIs from the NASA Exoplanet Archive \citep[][]{Akeson2013}, and the direct calibration of spectroscopic indices to empirically measured effective temperature and stellar radii, rather than using models to determine the effective temperature.  As with \citet{Mann2013b}, \citet{Mann2013c} does not include a metallicity correction for the radius-effective temperature relationship, which is predicted by nearly all evolutionary models.

In this paper we report new effective temperatures and metallicities for M Dwarf KOIs listed on the NASA Exoplanet Archive as of  August 2013, using $H$- and $K$-band spectroscopy.  When using spectroscopy, metallicity and temperature effects can be separated robustly by measuring equivalent widths of specific lines and the broad shape of the pseudo-continuum vs. wavelength \citep[e.g.][]{Rojas2010, Rojas2012, Terrien2012, Mann2013a}.  We use the indices and calibrations of \citet{Rojas2012} to determine stellar metallicity and effective temperature from the spectra.  We then interpolate those values onto new 5-Gyr isochrones from the Dartmouth Stellar Evolutionary Database to determine \mstar, \rstar, luminosity and distance.  The choice of age for the isochrones is  arbitrary, and does not significantly affect the resulting parameters due to the slow evolution of M dwarf stars after the pre-main sequence phase.  Nevertheless, we independently tested the age assumption using the publicly available TRILEGAL online interface \citep[e.g.][]{Girardi2005} and found a mean age for the M dwarf population in the Kepler field to be 3.2 Gyr ($M_\star < 0.5 M_\Sun$, $r$ magnitude $<$ 16.5).  Since there is no significant difference between the 3.2 Gyr predictions and 5 Gyr predictions for these particular stars, we proceed using the 5-Gyr predictions.  In this paper we only report on the stellar properties, leaving accurate transit fitting and planetary properties to a future publication.

In Section \ref{Survey} we discuss the sample of stars, the observations and the reduction of the data.  In Section \ref{Properties} we discuss the measurement of stellar properties and corresponding uncertainties, and compare our methods to other near-infrared methods for determining stellar properties.  In Section \ref{Comparison} we compare our results to \citet{Batalha2013}, \citet{Dressing2013}, \citet{Mann2013b} and \citet{Mann2013c} for our overlapping stars.  We find that our results generally agree with \citet{Dressing2013}, \citet{Mann2013b} and \citet{Mann2013c} but not \citet{Batalha2013}, which used values from the KIC.  Finally, in Section \ref{Conclusions}, we note several particularly interesting KOIs, summarize our results and reflect on the importance of obtaining accurate stellar properties for the full {\it Kepler} M dwarf target list.

\section{The Survey}\label{Survey}

\subsection{The Sample}\label{Sample}

We downloaded the list of KOIs available on the NASA Exoplanet Archive\footnote{\href{http://exoplanetarchive.ipac.caltech.edu/}{{http://exoplanetarchive.ipac.caltech.edu/}}} as of August 2013 without a ``false positive'' disposition as determined by the \ik pipeline, amounting to 3720 stars with 4636 transiting planet candidates \citep[NEA,][]{Akeson2013}.  From that list, we observed all KOIs with $r - J > 2.0$, where $r$ was measured in the KIC and $J$ was measured in 2MASS.  We use $r - J$ due to its sensitivity to stellar effective temperature and relative insensitivity to interstellar reddening, as well as the availability of both bands for nearly all of the KOIs (only 20 of the 3720 KOIs lack either band in the NEA).  \citet{Mann2012} used a cut of $K_P - J > 2.0$, where $K_P$ refers to the \ik magnitude, to successfully to isolate 98\% of M-type stars in a previous study concerning giant contamination in the {\it Kepler} sample \citep{Mann2012}.  We chose a similar approach, but use $r$ instead of $K_P$, since $K_P$ was not uniformly determined for the full Kepler target list \citep[see][]{Brown2011}.  Comparing the temperature estimates of \citet{Mann2013c} to $r$-$J$, we estimate this will include all dwarfs cooler than $\sim$4000 $K$.  In total, the sample amounts to 103 KOI stars.  

We note that three of the KOIs appear to be visual binary stars: KOI-1725, KOI-249 and KOI-4463 have nearby, bright companions within a few arcseconds.  We were unable to determine which object has the transit signal identified by {\it Kepler}, so we acquired spectra of both.  We report stellar properties for each of the stars within each of these systems, reporting the brighter component as `A' (e.g. KOI-249 A) and the fainter as `B' (e.g. KOI-249 B) as seen on the TripleSpec $K$-band guider camera.  KOI-4463 consists of two stars of near equal brightness, so we report them as KOI-4463 SE and NW for southeast and northwest.

We also include KOI-256 in our survey despite being identified as a false-positive in a previous paper \citep{Muirhead2013}.  KOI-256 is listed as a ``candidate'' on the NEA, but in fact the transit signal is a cool white dwarf passing behind the M dwarf primary: a post-common envelope system.  \citet{Muirhead2013} reported empirical properties for the M dwarf including high metallicity ([M/H] = +0.31 $\pm$ 0.10), suggesting possible pollution of the M dwarf atmosphere during an evolved state of the white dwarf progenitor.  The M dwarf is also rapidly rotating synchronously with the white dwarf orbit, and rapid rotation has been suggested as a mechanism for inflating M dwarf radii \citep{Chabrier2007}.  We therefore include KOI-256 in this study because the differences between the properties of the M dwarf in KOI-256 and the determinations using the $K$-band indices and Dartmouth Isochrones for the other KOIs shed light on the role of binary evolution and rapid rotation on stellar property determinations for M dwarfs.

\subsection{Observations}\label{Observations}

We observed the Cool KOIs with the TripleSpec Spectrograph at the Palomar Observatory 200-inch Hale Telescope.  TripleSpec is a near-infrared slit spectrograph covering 1.0 to 2.5 $\rm \mu m$ simultaneously with a resolving power $(\lambda/\Delta\lambda)$ of 2700 \citep{Wilson2004, Herter2008}.  Observations were carried out in June and August of 2011 for the KOIs from \citet{Borucki2011b}, with results published in a previous letter \citep{Muirhead2012b}.  In July and August of 2012, we augmented observations to include all of the Cool KOIs  in \citet{Batalha2013} using the sample definition above.  In July and August of 2013, we observed the remaining Cool KOIs listed on the NEA as of the observations.  All exposures used double-correlated sampling (equivalent to a ``Fowler depth'' of 1), which is appropriate given that all observations were photon-limited rather than read-noise limited \citep{Fowler1990}.  At the beginning of each night we observed dome darks and dome flats for flat-fielding.

Different observing strategies were used for the 2011 and 2012/13 observations.  For the 2011 observations, we used two positions on the slit, A and B, and exposures were taken in a ABBA pattern with 60 second integration times at each position.  Multiple ABBA sets were taken and combined until each reduced spectrum had a median per-channel signal-to-noise of at least 60.  

For the 2012/13 observations, we used four positions on the slit, ABCD, so as to reduce the effects of hot and dead pixels on the spectrograph detector.  Unlike in 2011, we used different exposure times ranging from 10 to 120 seconds per exposure, based on the signal-to-noise results from the 2011 observations.  A median signal-to-noise of at least 60 was achieved on all KOIs in the sample.

For each KOI observation, we also observed a rapidly rotating A0V star near in time and location in the sky to calibrate and remove absorption lines imprinted on the KOI spectrum by the Earth's atmosphere.  The A0V stars were chosen from a list of A0V stars returned by SIMBAD, with measured projected rotational velocities (\vsini) of greater than 100 \kms and no evidence of binarity.  For each KOI, the nearest A0V star with a $K$-band magnitude equal-to or less-than the KOI was chosen from the list and observed so as to achieve a signal-to-noise twice that achieved on the KOI.  Each A0V star observation was taken within 30 minutes of each KOI observation with an airmass difference of less than 0.1.

\subsection{Data Reduction}\label{Reduction}

For the 2011 observations, each A and B pair was differenced, flat-field-corrected and extracted using the SpexTool reduction package \citep[][updated for Palomar-TripleSpec]{Cushing2004}, which includes a program called Xtellcor for calibrating and removing the absorption lines introduced by the Earth's atmosphere using the respective observations of A0V stars \citep{Vacca2003}.  SpexTool incorporates procedures to remove detector effects such as quadrant-to-quadrant cross-talk and channel-to-channel capacitive coupling specific to the TripleSpec infrared detector \citep[artifacts and removal are described in][]{Muirhead2011}.  For the 2012 observations, the A and B observations and C and D observations were differenced separately, and extracted and combined using SpexTool.   Observations of the same KOI on different nights were combined using SpeXTool.

Absolute radial velocity differences between the KOIs can shift the absorption lines used to determine stellar properties.  To correct for this, we cross-correlated each spectrum with a spectrum of a similar spectral type from the IRTF Spectral Library \citep{Cushing2005, Rayner2009}, which have already been shifted to zero radial velocity and vacuum wavelengths.  We then apply an offset to the wavelength solution for the KOI spectra based on the result of the cross-correlation.  We visually inspected each KOI spectrum to ensure the locations of prominent absorption lines were consistent with those in the IRTF spectral library.  The spectra for the dwarf stars are plotted in Figure \ref{spectra}, and the spectra for the giant star KOI-977 and peculiar star KOI-3497 are plotted in Figure \ref{giant_spectra}.

\section{Stellar Properties}\label{Properties}

Dwarf/giant discrimination was performed on each star by visually inspecting the absorption lines in the CO band at 2.3 $\mu m$.  M giant stars have markedly deeper CO lines than dwarfs due to effects from low surface gravity \citep[e.g.][]{Rayner2009}.  We identify one giant star among the sample: KOI-977.  KOI-977 was previously identified as a giant in \citet{Muirhead2012b} using the same data in this paper, and later investigations by \citet{Mann2012} and \citet{Dressing2013} also identified KOI-977 as a giant star.  We note that in an adaptive optics imaging survey of KOIs, \citet{Law2013} found no obvious companions to KOI-977.  

For the remaining dwarf stars, stellar properties were determined using the calibrations of \citet{Rojas2012}: metallicity was determined using the equivalent widths of the $K$-band Na I doublet and Ca I triplet and corrected for temperature dependence using the $\rm H_2O-K2$ index.  \citet{Rojas2012} calibrated the relationship between equivalent width, $\rm H_2O-K2$ and metallicity using M dwarfs with F, G or K-type companions of known metallicity using the TripleSpec Spectrograph at Palomar: the very spectrograph used for the observations in this paper.  The calibration stars consisted of M dwarf stars with spectral types ranging from M1 to M6, with effective temperatures of between 2830 $K$ and 3850 $K$, metallicities ([Fe/H]) of between -0.69 and 0.24.  Stellar effective temperature was determined by interpolating metallicity and the $\rm H_2O-K2$ index onto a grid of synthetic temperature, metallicity and $\rm H_2O-K2$ indices calculated from the BT-Settl-2010 synthetic spectra assuming a surface gravity log(g) = 5.0 \citep{Allard2012}.  Unlike the metallicity determinations, the effective temperature determinations are therefore model-dependent and should be taken with some caution.

To estimate the uncertainties, we follow a Monte Carlo approach.  For each cool KOI spectrum, we create 100 copies of the spectrum, each with random Gaussian noise added to each channel based on the per-pixel uncertainties from photon noise, readout noise and flat-fielding errors for both the target data and the telluric calibrator data.  We then measure the $K$-band Na I doublet and Ca I triplet equivalent widths and $\rm H_2O-K2$ index for each of the 100 noisy copies.  We use calculate the resulting metallicities and effective temperatures for each of the noisy copies.  From these distributions, we calculate the standard deviation of the metallicities and effective temperatures, and use that as the uncertainty due to measurement errors.

Since the effective temperature determination incorporates the metallicity determination when interpolating onto the $\rm H_2O-K2$ indices calculated from the BT-Settl-2010 synthetic spectra, there is the possibility that the metallicity uncertainties and the effective temperature uncertainties are correlated.  The Monte Carlo approach enables a simple calculation of the Pearson correlation coefficient between these two quantities, using the 100 distributions of random effective temperature and metallicity measurements determined from the artificially noisy spectra.  Taking all of the cool KOIs in our sample, we found a mean correlation coefficient of -0.05, with the correlations coefficients producing a roughly normal distribution with a standard deviation of 0.1.  This indicates a low degree of correlation between these two parameters.

In addition to uncertainty due to measurement errors, there is also a systematic uncertainty due to any errors in the \citet{Rojas2012} calibration itself.  We assume a systematic uncertainty of 0.12 dex for the metallicity and 50 $K$ for the effective temperature, added in quadrature to the uncertainties from noise in the spectra.  The systematic metallicity uncertainty is based on the residuals between the \citet{Rojas2012} calibration and M dwarfs with known metallicities based on associated sun-like stars.  The systematic effective temperature uncertainty is an educated guess based on the few M dwarfs with both $K$-band effective temperatures and effective temperatures measured with optical long-baseline interferometery \citep[see Figure 3, panel a in ][]{Muirhead2012b}.

Several of the stars in the target sample have $\rm H_2O-K2$ values consistent with stars earlier than M0, where the \citet{Rojas2012} indices are not calibrated: KOIs 641, 1393, 2417, 2418, 2992, 3140, 3414, 4087 and 4463 SE and 4463 NW.  We include the spectra in this paper; however, we do not report stellar parameters for these stars as the \citet{Rojas2012} calibrations do not apply for stars earlier than M0.  Subtracting these stars and KOI 977 from the sample of 103 KOIs, this leaves 93 KOIs for which we report stellar properties, two of which are binary stars (KOIs 1725 and 249).

\subsection{Comparison of Metallicity Methods}

\begin{figure*}[h!]
\begin{center}
\includegraphics[width=7.0in]{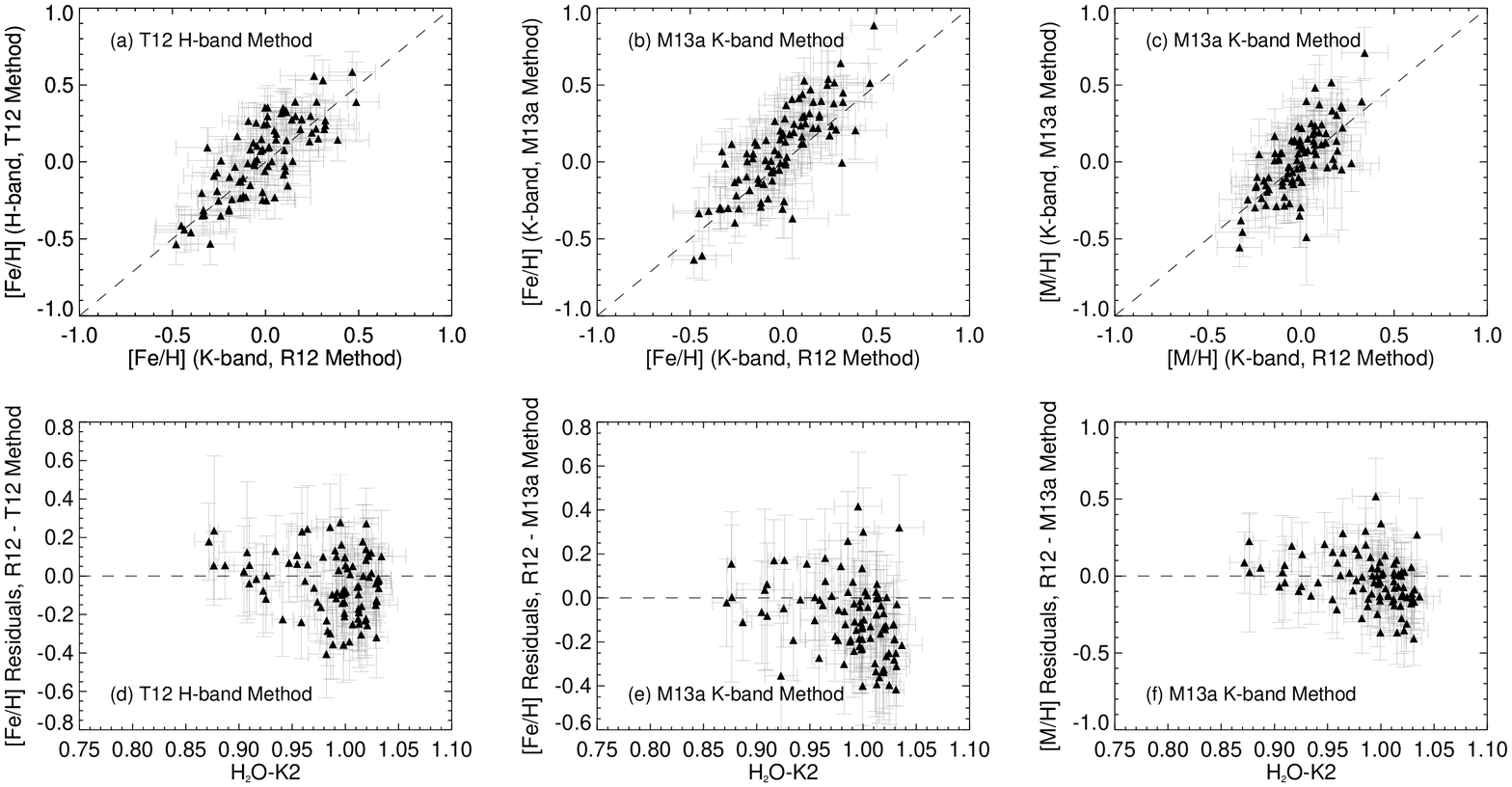}
\caption{Comparison of metallicity methods applied to the spectra of the KOIs in this survey. {\it Top Row}: Metallicity ([Fe/H] and [M/H]) as determined by the $H$-band method of \citet[][T12]{Terrien2012} and $K$-band method of \citet{Mann2013b} compared to the $K$-band method of \citet[][R12]{Rojas2012} on the KOIs observed in this paper.  {\it Bottom Row}: Residuals between methods as a function of the $\rm H_2O$-K2 index, which traces $T_{\rm eff}$.  The different methods clearly report different metallicities as a function of spectral type.  The discrepancies could be due to using different spectrographs (SpeX for T12 and M13b, and TripleSpec for R12), although the different resolution is accounted for and wavelength differences are removed.\label{metallicity_method_comparison}}
\end{center}
\end{figure*}

In addition to the work of \citet{Rojas2012}, there are two other calibrations relating near-infrared spectroscopic indices to M dwarf metallicity.   \citet{Terrien2012} describe a technique to measure M dwarf metallicities using the equivalent widths of Na I and Ca I lines in $H$-band and the $\rm H_2O-H$ index from \citet{Covey2010} for removing temperature effects.   They also developed their own $K$-band metallicity calibration using the same lines as \citet{Rojas2012}.  They show that their metallicity determinations are consistent with the \citet{Rojas2012} determinations for the M dwarfs that overlapped in their calibration samples.

In another paper, \citet{Mann2013a} explored indices at both optical and infrared wavelengths to find absorption lines that correlated strongest with metallicity for mid-K dwarfs through M dwarfs.  They found that the equivalent widths of the Na I doublet and CO (2-0) band head in $K$-band correlated strongest, and used the $\rm H_2O-K2$ index from \citet{Rojas2012} to remove temperature effects.  Unlike \citet{Rojas2012} and \citet{Terrien2012}, \citet{Mann2013a} included K dwarfs and stars with [Fe/H] $<$ -0.5 in their calibration sample.

For the sake of completeness, we calculate KOI metallicities using these two other techniques in addition to the \citet{Rojas2012} technique.  Unlike \citet{Rojas2012}, both \citet{Terrien2012} and \citet{Mann2013a} used the SpeX Spectrograph on the NASA Infrared Telescope Facility in Mauna Kea to acquire their spectra \citep{Rayner2003}.  SpeX has a resolving power of 2000, lower than TripleSpec's 2700, which may result in different equivalent width and temperature determinations.  We therefore convolve our TripleSpec spectra with a Gaussian kernel with a full-width-at-half-maximum matching a resolving power of 3000.  Convolving a resolution 2700 spectrum with a resolution 3000 kernel results in a spectrum with an equivalent resolution of 2000: the resolution of SpeX data before calculating the \citet{Terrien2012} and \citet{Mann2013a} metallicities.

Figure \ref{metallicity_method_comparison} plots the results using the \citet{Rojas2012} metallicity calibration vs. the results using the \citet{Terrien2012} and \citet{Mann2013a} calibrations.  We also plot the residuals vs. the H2O-K2 temperature indicator.  The methods are generally in agreement; however the largest scatter appears at high H2O-K2 values, or high temperatures.  The \citet{Mann2013a} determinations are occasionally unrealistically high, with several KOIs having [Fe/H] values greater than 0.5.  We speculate that the use of the CO (2-0) lines for metallicity determination is dangerous given its clear dependence on surface gravity.  However, it is possible that the high [Fe/H] values could be due to inaccurate accounting of the differences in resolution and wavelength calibration between the SpeX and TripleSpec spectra.  Either way, we choose to proceed using the \citet{Rojas2012} calibration, as it is most appropriate for the spectra we obtained.

\subsection{New Dartmouth Isochrones}

For those stars with $\rm H_2O-K2$ that do fall within the \citet{Rojas2012} calibration, stellar mass, radius and luminosity for the dwarfs were determined by interpolating the measured effective temperature and metallicity onto new 5-Gyr stellar evolutionary isochrones.  Stellar evolution models used in this study were computed using an updated version of the Dartmouth series. Of primary significance for this work is that the fitting point of the surface boundary conditions has been modified. Models presented by \citet{Dotter2008} determine the initial pressure and temperature of the model envelope integration by 
finding the gas pressure of a PHOENIX model atmosphere structure at T = \teff for a given logg and [Fe/H].  However, for this work, the models fit the surface boundary conditions at an larger optical depth. For a given Teff, logg, and [Fe/H], the starting pressure and temperature for the envelope integration are defined using the same model atmosphere structures, but at and optical depth of $\tau = 10$. Non-adiabatic effects in the outer layers of M dwarfs are, as a result, treated more realistically \citep{Chabrier1997}.

This has two consequences for the Dartmouth models. First, stellar models 
are more accurate below 0.2 $M_\Sun$. The original Dartmouth series are known 
to be inaccurate below 0.2 $M_\Sun$, a problem stemming from the adopted 
treatment of the surface boundary conditions. A second consequence is that 
the models have a revised mass-Teff relation, whereby models of a given 
mass have a hotter Teff in the updated models. This is most pronounced 
below \teff $\sim$ 4700 K, with the largest difference of about 60 K 
occurring near the fully convective boundary. Radii, however, are relatively 
unaffected.

Models were computed for masses above 0.08 Msun for metallicities 
-0.5 $<$ [Fe/H] $<$ +0.5. Grid spacings were 0.05 Msun in mass and 0.1 dex
in metallicity. A finer mass resolution of 0.01 Msun along the isochrones 
was then acheived by cubic-spline interpolation. This fine grid allow for
more accurate interpolation of empirical data onto stellar isochrones.

We interpolated the measured effective temperature and metallicity onto the isochrones by first constructing a Delaunay triangulation from the effective temperatures and metallicities listed in the isochrones using the ``triangulate'' procedure in IDL \citep[e.g.][]{Lee1980}.  We then interpolated the stellar mass, radius, luminosity and expected absolute $K$-band for each target by interpolating the measured values onto the 3-dimensional temperature-metallicity-mass, temperature-metallicity-radius, etc, plane using the ``trigrid'' procedure in IDL.

Once settled onto the main-sequence (age $\gtrsim$ 0.5 to 1.0 Gyr), M dwarfs do not significantly evolve for tens of billions of years \citep{Laughlin1997}.  Therefore, unless a particular object is a pre-main-sequence star, the choice of 5-Gyr isochrones will not strongly affect the resulting masses and radii.  The isochrones include predictions for the absolute 2MASS $K$-band magnitudes, based on the PHOENIX model atmospheres used to set the surface boundary conditions in the evolutionary models \citep{Dotter2008, Hauschildt1999a, Hauschildt1999b}.  For each KOI we compare the predicted absolute $K$-band magnitude from the isochrones to the apparent $K$-band magnitude as measured by 2MASS to estimate the distance to the KOI.  Using this technique, we measure a median distance to the stars in the sample to be 131 pc, with the maximum distance being 622 pc for KOI-2926.  The stars are significantly closer than the four open clusters in the \ik field of view: NGC 6791, NGC 6811, NGC 6819 and NGC 6866, which have distances of 4100, 1215, 2360 and 1450 parsecs respectively, and $A_V$ extinction factors of 0.117, 0.160, 0.238 and 0.169 respectively, as listed in the WEBDA database of clusters \citep{Mermilliod1995}.  Since $A_K$ is an order of magnitude less than $A_V$, we expect extinction to cause effects at less than the 0.01 magnitude level, which is near the uncertainty on the measured 2MASS magnitudes.  Therefore, we do not correct for extinction when calculating the distances to the stars in our sample.

\section{Comparison to Published Properties}\label{Comparison}

In this section, we compare our measured stellar properties to those of other studies for overlapping Cool KOIs, specifically \citet{Batalha2013}, \citet{Dressing2013}, \citet{Mann2013b} and \citet{Mann2013c}.

\subsection{\citet{Batalha2013}}

\begin{figure*}[t]
\begin{center}
\includegraphics[width=7.0in]{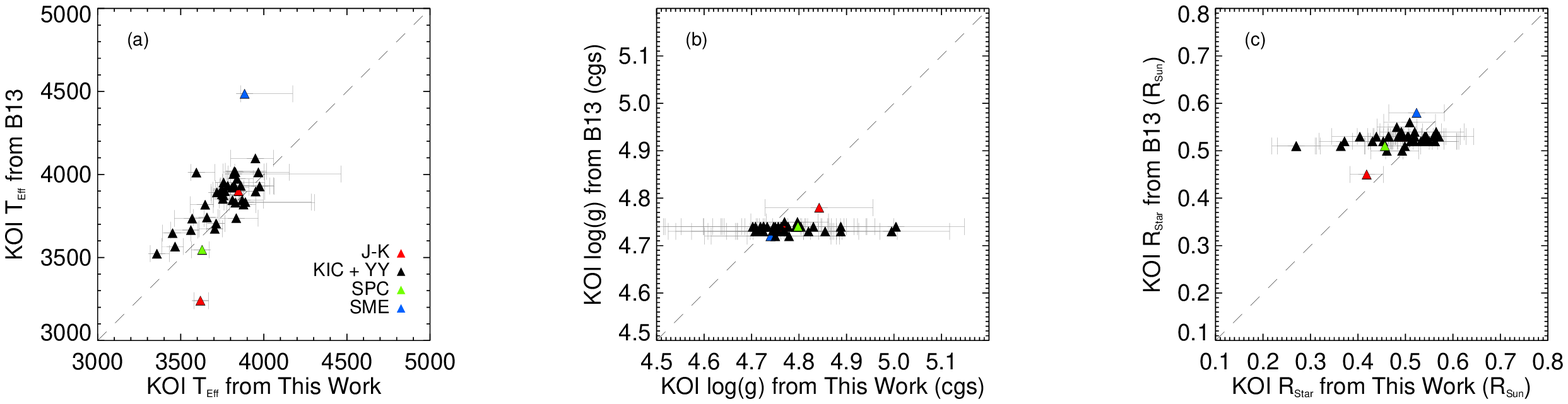}
\caption{Comparison between the stellar properties ascribed to the Cool KOIs in \citet[][B13]{Batalha2013} and this work.  {\it Panel a}: \teff from \citet[][]{Batalha2013} versus this work.  {\it Panel b}: log(g) from \citet[][]{Batalha2013} versus this work.  {\it Panel c}: $R_\star$ from \citet[][]{Batalha2013} versus this work.  Although the \teff determinations are relatively consistent, the use of Yonsei-Yale isochrones for determining properties of low-mass stars by \citet{Batalha2013} leads to very different \logg and \rstar determinations when compared to this work, which uses the Dartmouth isochrones \citep{Dotter2008}.  \citet{Boyajian2012} showed that the Dartmouth isochrones reproduce empirical radius determinations of low-mass stars more reliably than the Yonsei-Yale isochrones. \label{batalha_comparison}}
\end{center}
\end{figure*}

Figure \ref{batalha_comparison} compares the \teff, \logg and \rstar determinations from \citet{Batalha2013} to this paper, for KOIs that overlap.  \citet{Batalha2013} used four different methods for determining stellar effective temperature, and used the Yonsei-Yale isochrones to interpolate \rstar and \mstar \citep{Yi:2003fu}.  In the figure, ``J-K'' refers to the use of $J$-$K$ colors to determine \teff.  ``KIC + YY'' refers to the use of the KIC derived \teff  and \logg as starting points for a parameter search using the Yonsei-Yale isochrones \citep{Yi:2003fu}.  ``SPC'' and ``SME'' refer to spectral synthesis codes used in coordination with the Yonsei-Yale isochrones to determine properties for those KOIs with optical spectroscopy \citep{Buchhave2012,Valenti1996,Valenti2005}.

The Yonsei-Yale isochrones reproduce properties for Sun-like stars reasonably well.  However, for low-mass stars they predict a \rstar-\teff relationship that flattens out around 0.5 \rstar for \teff at or below 4000 K.  Empirical measurements of low-mass star radii using optical long-baseline interferometry do not support this relationship, instead favoring the predictions of the Dartmouth isochrones \citep{Boyajian2012} and BCAH isochrones \citep{Baraffe1998}, which predict a flatting out of stellar radius closer to 3000 $K$ and 0.1 \rstar due to effects from electron degeneracy pressure.    The difference in choice of isochrones leads to a dramatic difference in the stellar radius determinations between this work and \citet{Batalha2013}, despite relative agreement between our \teff determinations.  We note, however, that recently revised Yonsei-Yale isochrones correct for these discrepancies \citep{Spada2013}, but the isochrones used in \citet{Batalha2013} did not.

\subsection{\citet{Dressing2013}}

\begin{figure*}[t]
\begin{center}
\includegraphics[width=7.0in]{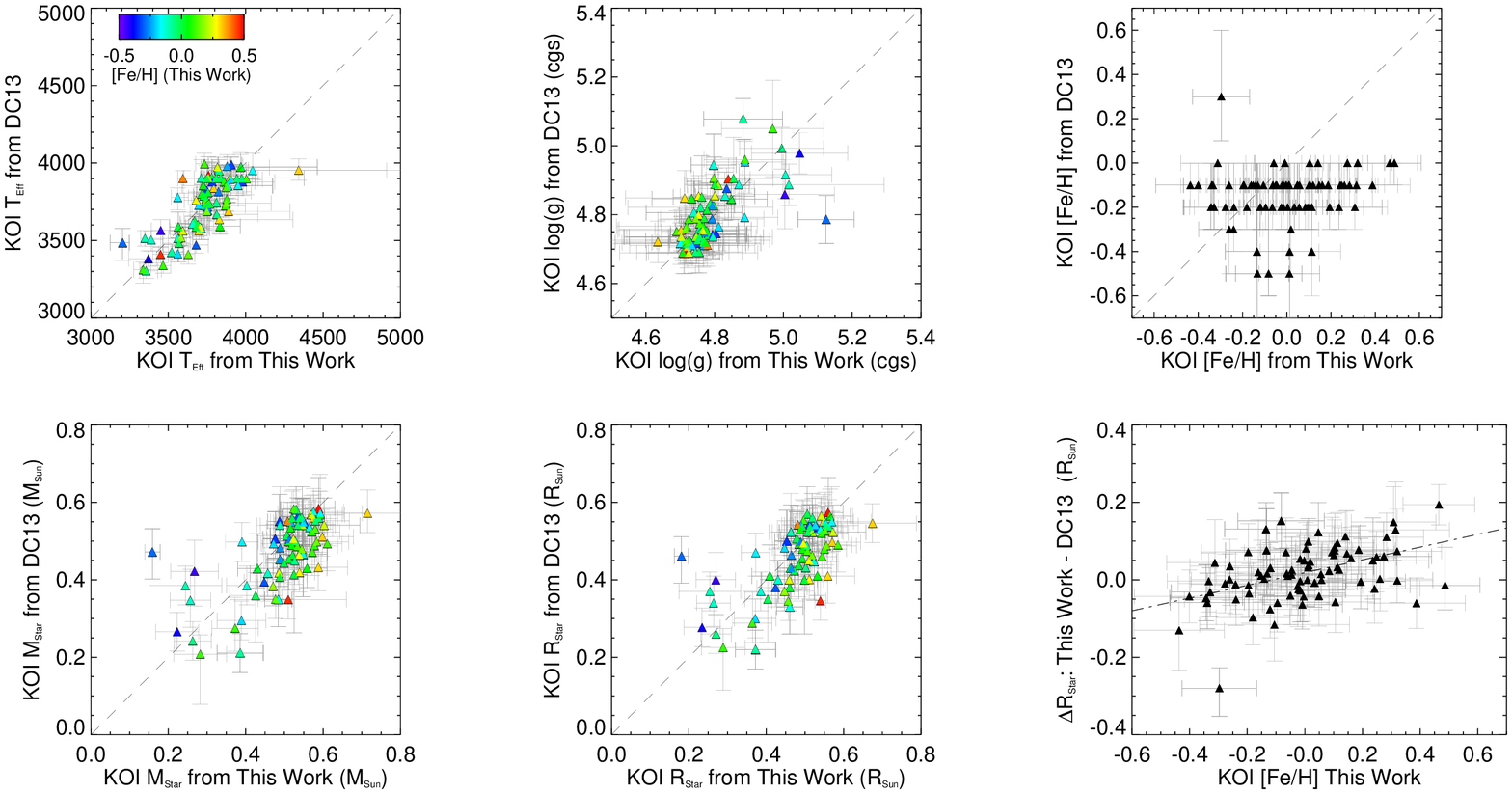}
\caption{Comparison between the stellar properties ascribed to the Cool KOIs in \citet[][D13]{Dressing2013} and this work for stars that overlap in both studies.  {\it Top row}: \teff, \logg and [Fe/H] from \citet[][]{Dressing2013} versus this work.  In general, our \teff and \logg determinations agree, but the [Fe/H] determinations are admittedly untrustworthy in D13 owing to the challenge of determining metallicities with colors alone.  For the \teff and \logg plots we colored the symbols according to the [Fe/H] determinations from this work.  {\it Bottom row}: \mstar and \rstar from D13 compared to this work, and finally the difference between our \rstar determinations vs. [Fe/H] from this work.  We fit a line showing a slight, but apparent metallicity dependence to the residuals.\label{dressing_comparison}}
\end{center}
\end{figure*}

\citet{Dressing2013} used KIC and 2MASS photometry to determine stellar \teff, \rstar, \mstar, \logg, \feh  and distances for 3897 Cool \ik targets observed as part of the primary \ik planet search program.  Of the 93 dwarf Cool KOIs we observed with TripleSpec in this paper, 84 of them also have stellar properties reported in \citet{Dressing2013}.  The remaining 15 are stars that were added to the \ik target list via the \ik Guest Observer program, which is also surveyed for transiting planets in addition to the primary \ik exoplanet planet search program.

Figure \ref{dressing_comparison} compares the stellar parameter determinations of this paper to \citet{Dressing2013} for the 84 overlapping stars.  In general, the spectroscopic and photometric methods are in agreement, especially for \teff.  However, \citet{Dressing2013} had admittedly poor constraints on stellar metallicities, due to degeneracies between \teff and metallicity, which is reflected in panel c of the figure.  This leads to a slight metallicity dependence to the difference between our respective radius determinations, as shown in panel f.

\subsection{\citet{Mann2013b}}

\begin{figure*}[t]
\begin{center}
\includegraphics[width=7.0in]{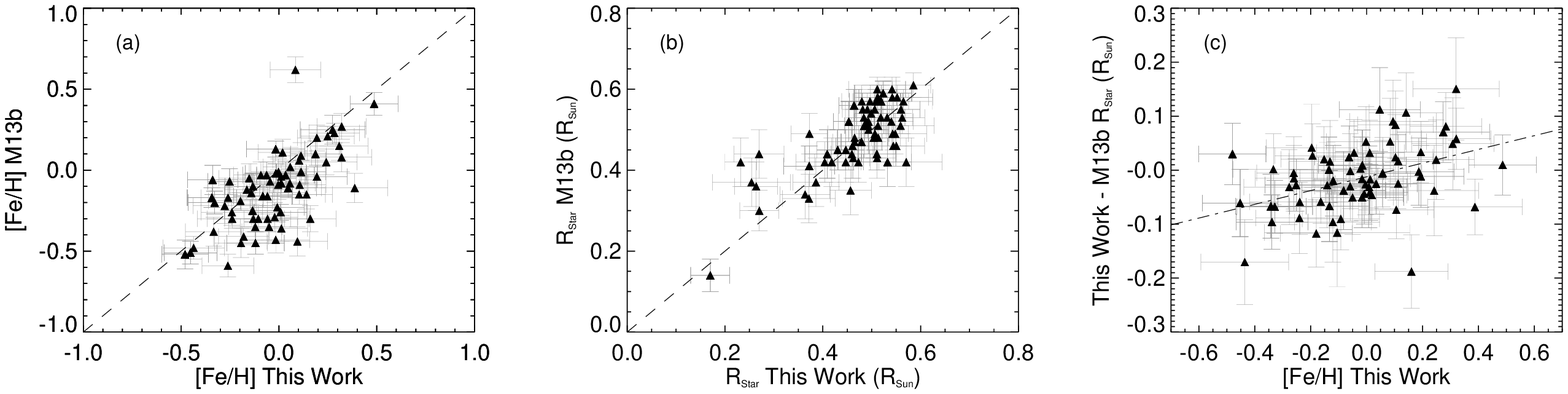}
\caption{Comparison between the stellar properties ascribed to the Cool KOIs in \citet[][M13b]{Mann2013b} and this work for stars that overlap in both studies.  {\it Panel a}: [Fe/H] from \citet[][]{Mann2013b} versus this work.  {\it Panel b}: \rstar from \citet[][M13b]{Mann2013b} compared to this work.  {\it Panel c}: The difference between our \rstar determinations vs. [Fe/H] from this work.  As with \citet{Dressing2013} we fit a line showing a slight, but apparent metallicity dependence to the residuals.\label{mann2013b_comparison}}
\end{center}
\end{figure*}

\citet{Mann2013b} acquired optical spectra of 67 of the 93 Cool KOIs in this sample.  In their paper they reported metallicities using the optical calibrations of \citet{Mann2013a}, and calculated effective temperatures for the stars using custom comparison to model stellar atmospheres.  They proceeded to estimate the radii of the M dwarfs using the \teff-\rstar relation from \citet{Boyajian2012}.  The \teff-\rstar relation in \citet{Boyajian2012} was fit using observations of M dwarf radii using optical long-baseline interferometry, where the authors did not detect a dependence on stellar metallicity.  However, both the Dartmouth and BCAH evolutionary isochrones predict a significant metallicity dependence to the  \teff-\rstar relation for main-sequence low-mass stars.  Figure \ref{mann2013b_comparison} compares the stellar parameter determinations of this paper to \citet{Mann2013b} for the 67 overlapping stars.

\subsection{\citet{Mann2013c}}

\begin{figure*}[t]
\begin{center}
\includegraphics[width=7.0in]{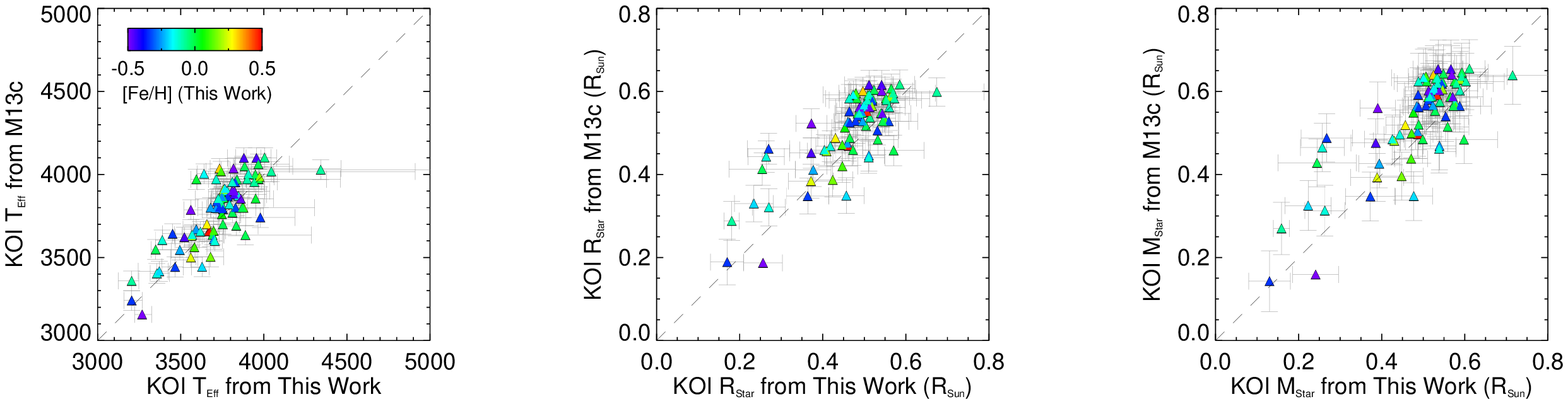}
\caption{Comparison between the stellar properties ascribed to the Cool KOIs in \citet[][M13c]{Mann2013c} and this work for stars that overlap in both studies.  {\it Left}: [Fe/H] from \citet[][]{Mann2013b} versus this work.  {\it Center}: \rstar from \citet[][M13c]{Mann2013c} compared to this work.  {\it Right}: The difference between our \rstar determinations vs. [Fe/H] from this work.  As with \citet{Dressing2013} we fit a line showing a slight, but apparent metallicity dependence to the residuals.\label{mann2013c_comparison}}
\end{center}
\end{figure*}

\citet{Mann2013c} acquire optical spectra of 81 of the 93 Cool KOIs in this sample.  They reported effective temperatures, stellar radii and masses by calibrating spectral indices directly to the {\it empirical} effective temperature and radius measurements of nearby M dwarfs from \citet{Boyajian2012}, and masses using mass-magnitude relationships from \citet{Delfosse2000}, without invoking atmospheric models as with \citet{Mann2013b}.  Figure \ref{mann2013c_comparison} compares the stellar parameter determinations of this paper to \citet{Mann2013c} for the 81 overlapping stars.

\subsection{Summary of Comparisons}

Despite the differences in methods, as with \citet{Dressing2013}, the stellar properties from \citet{Mann2013b} and \citet{Mann2013c} and our results are generally in agreement.  Figure \ref{mann2013b_comparison} compares the stellar parameter determinations of this paper to \citet{Mann2013b} for the overlapping stars.  The only striking difference is in the handful of stars with radii between 0.20 and 0.35 $R_{\rm Sun}$ as determined by our methods (panel b).  For those stars, \citet{Mann2013b} report significantly larger radii.  The \citet{Boyajian2012} relationship is sparsely sampled in this regime, leading to significant differences between the \citet{Boyajian2012} \teff-\rstar relation and predictions from the Dartmouth models at these particular radii.

\section{Interesting Systems}

\subsection{The Smallest KOIs}

We identify the following KOIs as M4V-dwarfs, making them the smallest stars with planet-candidates detected by Kepler: KOI-4290, KOI-2842, KOI-961 \citep[confirmed as \ik 42,][]{Muirhead2012a}, KOI-1725 B and KOI-2704.  In this paper, we choose to concentrate on the stars rather than their planets, we note that 3 of these 5 stars have multiple planet-candidates listed in the NEA as detected by the \ik Pipeline.

\subsection{KOI-3497 and KOI-1880}

KOI-3497 presents an interesting case.  At first, we classified KOI-3497 as a giant star due to the relatively deep CO lines in $K$ band.  However, the Na and Ca lines are more consistent with a dwarf star, and KOI-3497's proper motion from the UCAC4 catalog is consistent with being within 100 pc \citep[$\mu = 23.3 mas\,yr^{-1}$,][]{Zacharias2012}, supporting a nearby dwarf rather than distant giant.

We observed KOI-3497 with the Robo-AO laser adaptive optics and imaging system \citep{2013JVE....7250021B} on the Palomar Observatory 60-inch Telescope on UT 2013 15 August to look for contaminating sources within the \textit{Kepler} aperture. We used a long-pass filter with a 600nm cut-on (LP600) to more closely approximate the \text{Kepler} bandpass while maintaining diffraction-limited resolution \citep{Law2013}. The observation consisted of a sequence of full-frame-transfer detector readouts at the maximum rate of 8.6 Hz for a total of 90 s of integration time. The individual images were then combined using post-facto shift-and-add processing using KOI-3497 as the tip-tilt star with 100\% frame selection. We detected that KOI-3497 is a binary with a separation of ${0}\farcs{82} \pm {0}\farcs{06}$; the secondary component is at a position angle of $176 \pm 3$ degrees with respect to the primary.

\begin{figure}[t]
\begin{center}
\includegraphics[width=2.0in]{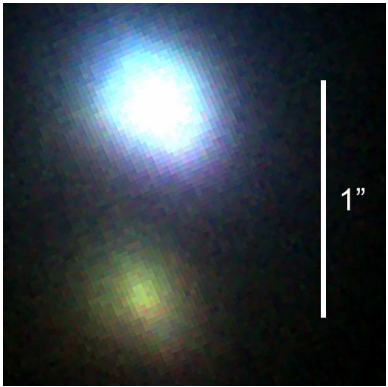}
\caption{False color Robo-AO image of KOI-3497. Sloan $rÕ$, $iÕ$ and $zÕ$ band images have each been normalized to 25\% of their primary peak intensity and have been assigned to B, G and R color channels respectively.  The scaling is linear, and North is up, East is left.\label{roboao}}
\end{center}
\end{figure}

We acquired additional observations with Robo-AO on UT 2013 24 October to determine the relative color of each object (see Fig. \ref{roboao}). Integrations of 120 s in each of the LP600 and Sloan $r'$-, $i'$- and $z'$-band filters \citep{York2000} were captured. We used simple aperture photometry to calculate the flux ratio of the primary and secondary components: we measured the total flux centered on each component and subtracted off an equivalent aperture on the opposite side of the companion to approximate subtraction of each stellar halo and sky background. Multiple apertures sizes were used to generate estimates of the systematic errors, and we found consistent flux ratios when using apertures from {0}\farcs{17} to {0}\farcs{47}. We found the following magnitude differences, $\Delta m_{LP600} = 1.47 \pm 0.12$, $\Delta m_{r'} = 2.10 \pm 0.16$, $\Delta m_{i'} = 1.42 \pm 0.27$, $\Delta m_{z'} = 1.13 \pm 0.23$, indicating that the secondary is redder than the primary.

Without individual spectra of the two objects in KOI-3497, it is difficult to assign reliable effective temperatures for the two objects.  However, in a study of low-mass stars in the Sloan Digital Sky Survey,  \citet{Bochanski2007} provided useful $r-i$ and $i-z$ color-spectral type relations for M type stars (their Table 1).  Combining the calibrated SDSS-like photometry from the KIC (providing $g$, $r$ and $i$) with the Robo-AO colors, we determine the following $r-i$ colors for KOI-3497: 0.278 for the brighter component and 0.958 for the fainter.  Following \citet{Bochanski2007}, the brighter component is bluer than an M0 star, however the fainter component is consistent with either an M2 or M3 star.  Therefore, we label the object as peculiar, and attribute the deep CO relative to Na and Ca as a combination of a K-dwarf and M-dwarf star.

We also note that a recent paper by \citet{Law2013} indicates that KOI-1880 also contains a fainter contaminant  1.7" from the brighter star.  However, the magnitude difference in LP600 is 3.7, meaning the infrared spectra are likely not significantly contaminated.

\section{Conclusions}\label{Conclusions}

Table \ref{format_table} lists the stellar properties that are included in the machine-readable table (MRT) available in the online version of the journal.  We also include the $H$- and $K$-band spectra for download.  The spectra presented in this paper represent a significant investment of observational resources.  However, the spectra and stellar properties direct future observational resources invested in the Cool KOIs towards high-priority targets.  We note several mid-M dwarf KOIs: KOI-4290, KOI-2842, KOI-961 (now \ik 42), KOI-1725 B and KOI-2704.  KOI-2842 and KOI-2704 both appear to host multiple short-period transiting planet candidates; however, follow up transit analysis is necessary to determine their true sizes and confirm that they are {\it bona fide} exoplanet hosts.

\begin{figure*}[t]
\begin{center}
\includegraphics[width=5.0in]{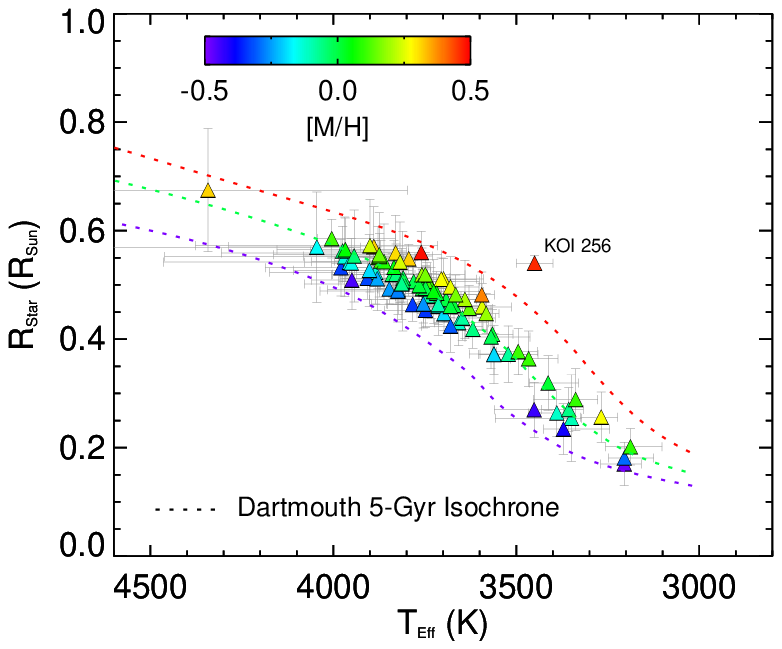}
\caption{\teff, \mh and \rstar determinations for the Cool KOIs based on this work.  For most of the Cool KOIs, we measure \teff and \mh using the $K$ band spectral indices of \citet{Rojas2012}, then interpolate those onto 5-Gyr ischrones from the Dartmouth Stellar Evolution Database \citep[{\it dashed lines}][]{Dotter2008}.  For KOI-256 (labelled) and KOI-961 (now \ik 42), we determined the stellar properties using markedly different techniques, with less influence from predictions of isochrones \citet[][respectively]{Muirhead2012a, Muirhead2013}.  The location of KOI-256 indicates that the Dartmouth Isochrones may be underpredicting stellar radii at a given \teff and \mh.\label{teff_vs_rad}}
\end{center}
\end{figure*}

Figure \ref{teff_vs_rad} plots \rstar vs. \teff for the stars observed in this survey including the isochrone predictions used in this study.  The alignment along the predictions is by construction.  We also include KOI-256, which has a \teff and \mh determined from a $K$ band spectrum like the rest of the KOIs, but the radius was determined empirically using the eclipses and occultations of a white dwarf companion.  The empirically-determined radius is significantly higher than the predictions of the isochrones.  For one, KOI-256 is rapidly rotating with a period equal to that of the white dwarf orbit: 1.37865 $\pm$ 0.00001 days.  There is a long-standing discrepancy between the mass-radius relationships for M dwarf predicted by models, and those measured from eclipsing binary stars \citep[e.g.][]{Torres2002}.  One possible explanation for the discrepancy involves changes in the efficiency of convection with rapid rotation and magnetic fields \citep[e.g.][]{Cox1981, Mullan2001, Chabrier2007, MacDonald2012, MacDonald2013, Feiden2013}.  Eclipsing binary typically have short orbital periods due to selection effects, and the stars are believed to be synchronously rotating, including KOI-256 which shows rotational modulation matching the system orbital period \citep{Muirhead2013}.

The mass-radius discrepancy also applies to effective temperature-radius relationships.  In fact, \citet{Boyajian2012} found markedly different \teff-\rstar relationships for eclipsing binary M dwarfs and field M dwarfs, although the methods used to determine the effective temperatures are very different in both cases.  KOI-256 appears to be consistent with the eclipsing binaries.  Therefore, we do not believe the location of KOI-256 on Figure \ref{teff_vs_rad} indicates a substantial fault with the isochrone predictions.  Also, the KOI-256 system likely went through a common-envelope phase when the white dwarf progenitor was an evolved star.  The common-envelope phase could also introduce \teff-\rstar effects that the isochrones do not include.

\begin{table}
\begin{center}
\caption{Cool KOI Stellar Properties in the Online Machine-Readable Table\label{format_table}} 
\begin{tabular}{l c c r } 
\hline\hline                        
\\[-1.5ex]
Parameter & Units & MRT Format Code & Note \\ [1.0ex] 
\hline                  
\\[-1.5ex]
KOI Number &  & A7 & \\ [1.0ex]
Star KIC ID &  & L8 & \\ [1.0ex]
$T_{\rm Eff}$ & $K$ & I4 & 1 \\ [1.0ex]
$+\sigma_{T_{\rm Eff}}$ & $K$ & I4 & 1 \\ [1.0ex]
$-\sigma_{T_{\rm Eff}}$ & $K$ & I4 & 1 \\ [1.0ex]
$\rm [Fe/H]_{R12}$ & log(Solar) & F5.2 & 1 \\ [1.0ex]
$\sigma_{\rm [Fe/H]_{R12}}$ & log(Solar) & F4.2 & 1 \\ [1.0ex]
$\rm [M/H]_{R12}$ & log(Solar) & F5.2 & 1 \\ [1.0ex]
$\sigma_{\rm [M/H]_{R12}}$ & log(Solar) & F4.2 & 1 \\ [1.0ex]
Sp Type & KHM & A4 & 1 \\ [1.0ex]
$M_\star$ & $M_\Sun$ & F4.2 & 2 \\ [1.0ex]
$\sigma_{M_\star}$ & $M_\Sun$ & F4.2 & 2 \\ [1.0ex]
$R_\star$ & $R_\Sun$ & F4.2 & 2 \\ [1.0ex]
$\sigma_{R_\star}$ & $R_\Sun$ & F4.2 & 2 \\ [1.0ex]
Distance & pc & I3 & 3 \\ [1.0ex]

    \hline 
    \\[-1.5ex]

$\rm [Fe/H]_{T12}$ & log(Solar) & F5.2 & 6 \\ [1.0ex]
$\sigma_{\rm [Fe/H]_{T12}}$ & log(Solar) & F4.2 & 6 \\ [1.0ex]

$\rm [Fe/H]_{M13}$ & log(Solar) & F5.2 & 7 \\ [1.0ex]
$\sigma_{\rm [Fe/H]_{M13}}$ & log(Solar) & F4.2 & 7 \\ [1.0ex]
$\rm [M/H]_{M13}$ & log(Solar) & F5.2 & 7 \\ [1.0ex]
$\sigma_{\rm [M/H]_{M13}}$ & log(Solar) & F4.2 & 7 \\ [1.0ex]

\tablenotetext{1}{Determined using the $K$ band spectroscopic indices of \citet{Rojas2012}.  KHM refers to the \citet{Kirkpatrick1991} spectral type system.  For KOI 961, we use the effective temperature from \citep{Muirhead2012a}, which was determined by comparison to empirical measurements of Barnard's Star.}
\tablenotetext{2}{stellar properties determined by interpolation onto a 5-Gyr isochrone from the Dartmouth Stellar Evolution Database \citep{Dotter2008}.}
\tablenotetext{3}{Distance calculated by combining the predicted $M_K$ from the Dartmouth isochrones to the apparent $K$ band magnitude from 2MASS \citep{Cutri2003}.}
\tablenotetext{4}{For completeness, we include metallicity determinations using $H$-band indices of \citep{Terrien2012}, referred to as T12, and the $K$ band indices of \citet{Mann2013a}, referred to as M13.}
\end{tabular}
\end{center}
\end{table}

\begin{figure*}[]
\begin{center}
\includegraphics[width=6.5in]{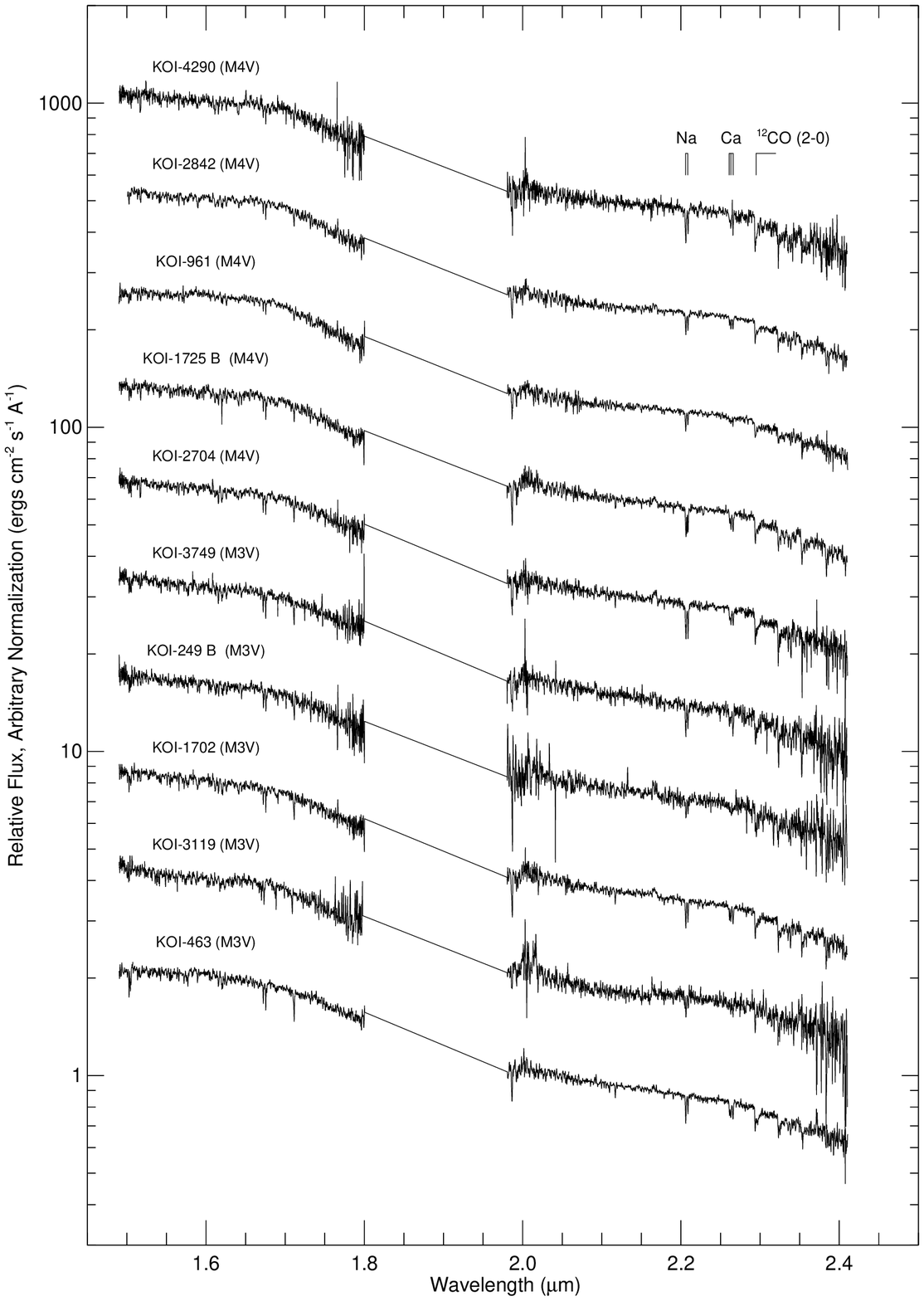}
\caption{\spectra\label{spectra}}
\end{center}
\end{figure*}

\addtocounter{figure}{-1}
\begin{figure*}[]
\begin{center}
\includegraphics[width=6.5in]{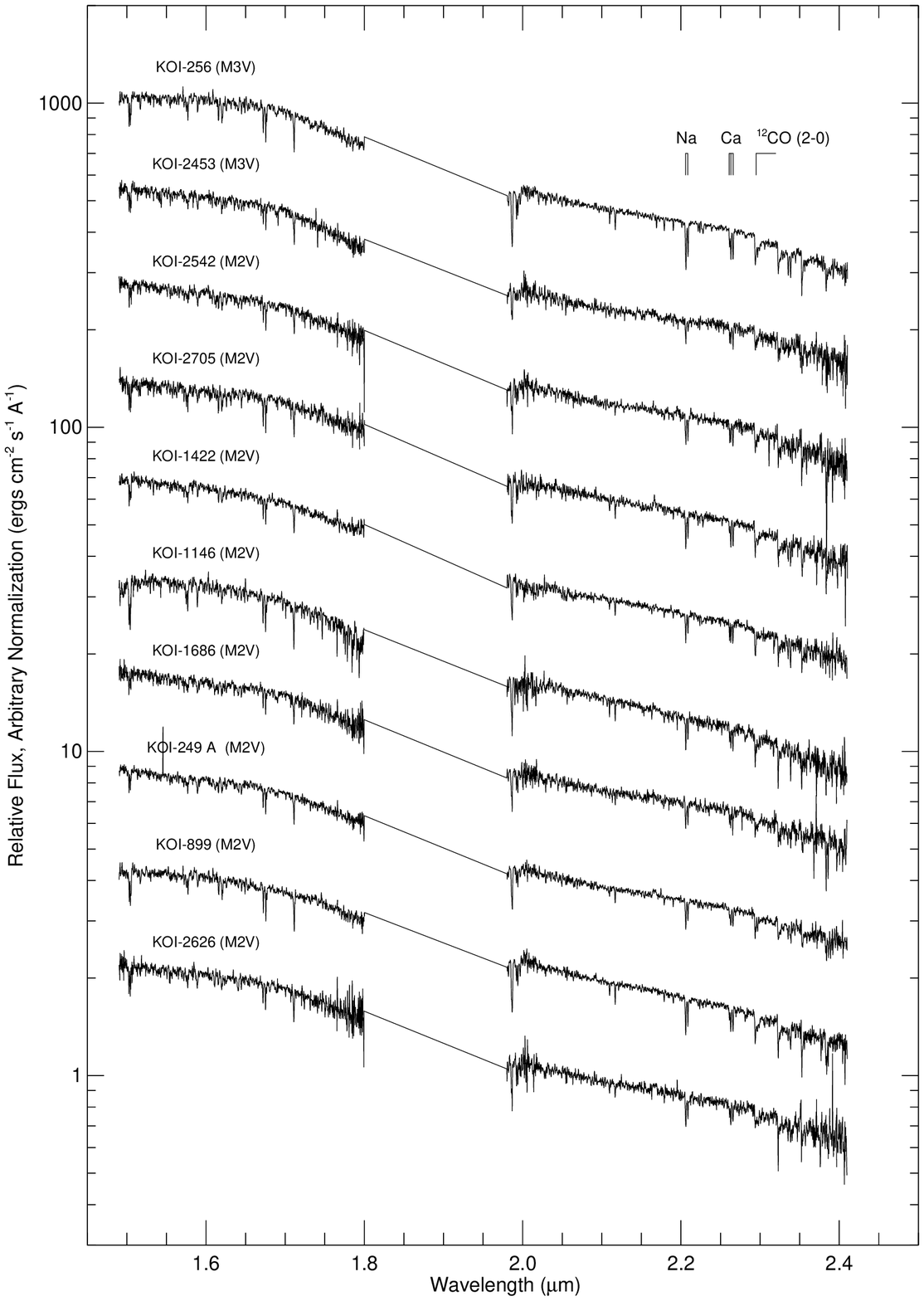}
\caption{continued.  \spectra}
\end{center}
\end{figure*}

\addtocounter{figure}{-1}
\begin{figure*}[]
\begin{center}
\includegraphics[width=6.5in]{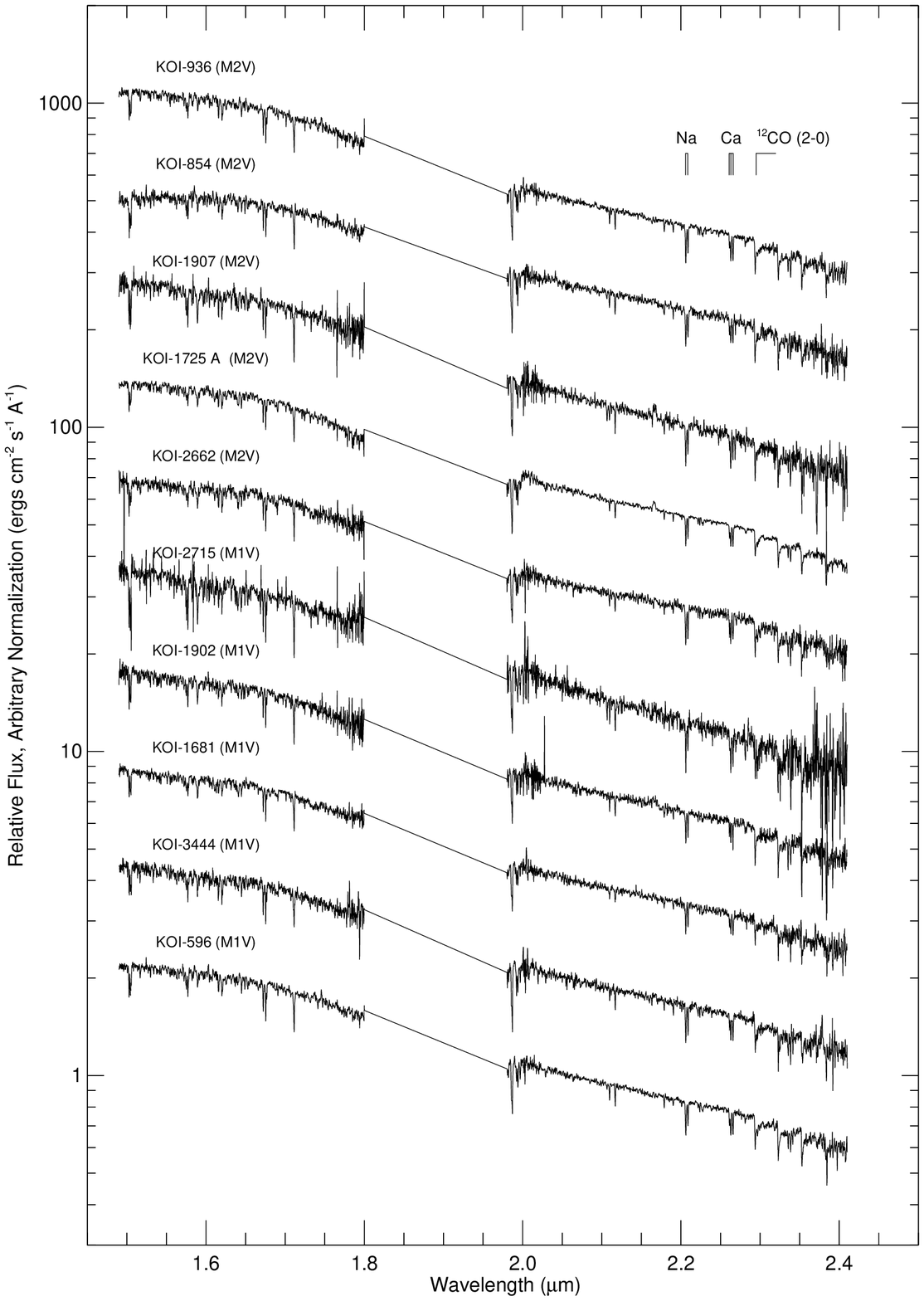}
\caption{continued.  \spectra}
\end{center}
\end{figure*}

\addtocounter{figure}{-1}
\begin{figure*}[]
\begin{center}
\includegraphics[width=6.5in]{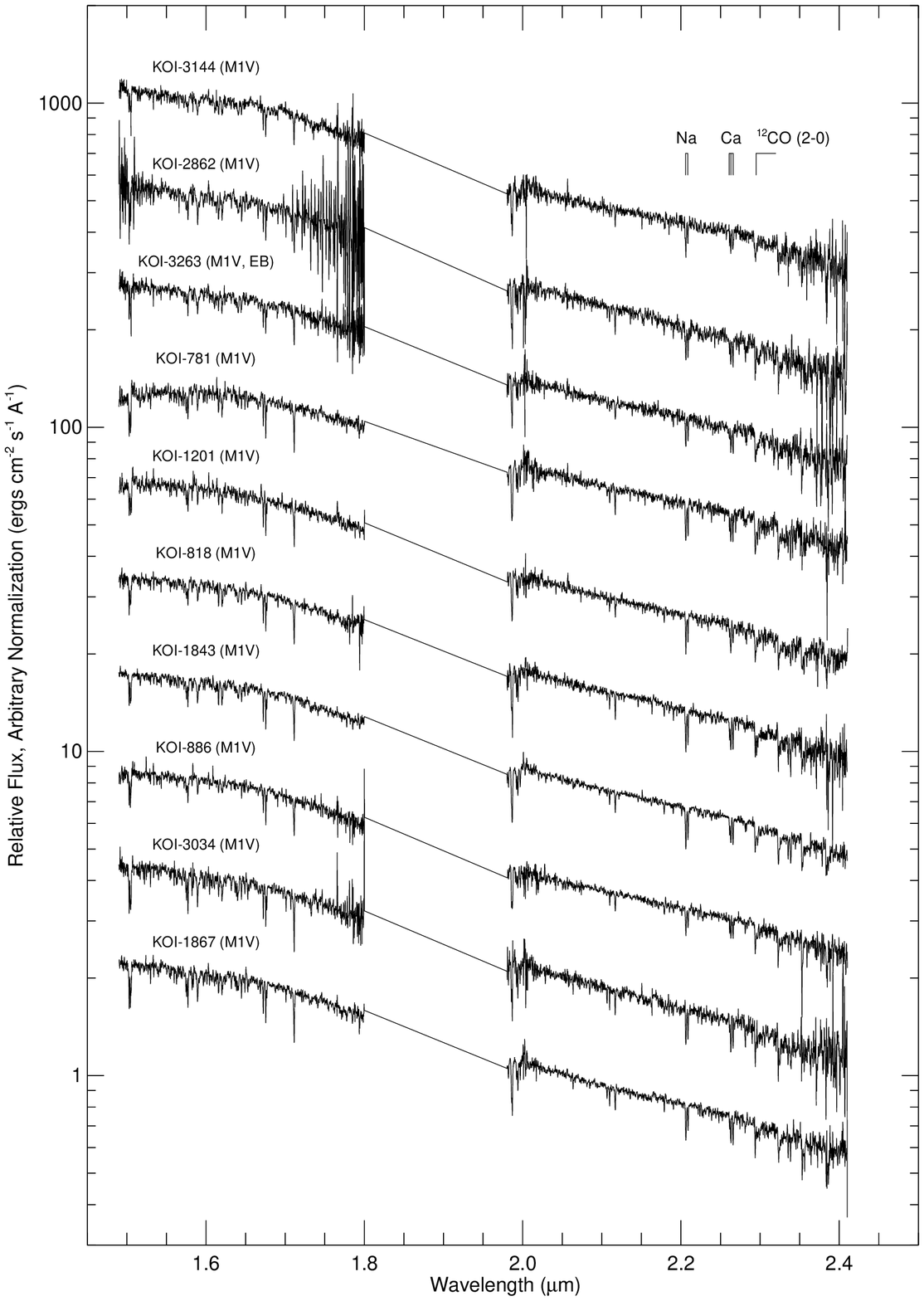}
\caption{continued.  \spectra}
\end{center}
\end{figure*}

\addtocounter{figure}{-1}
\begin{figure*}[]
\begin{center}
\includegraphics[width=6.5in]{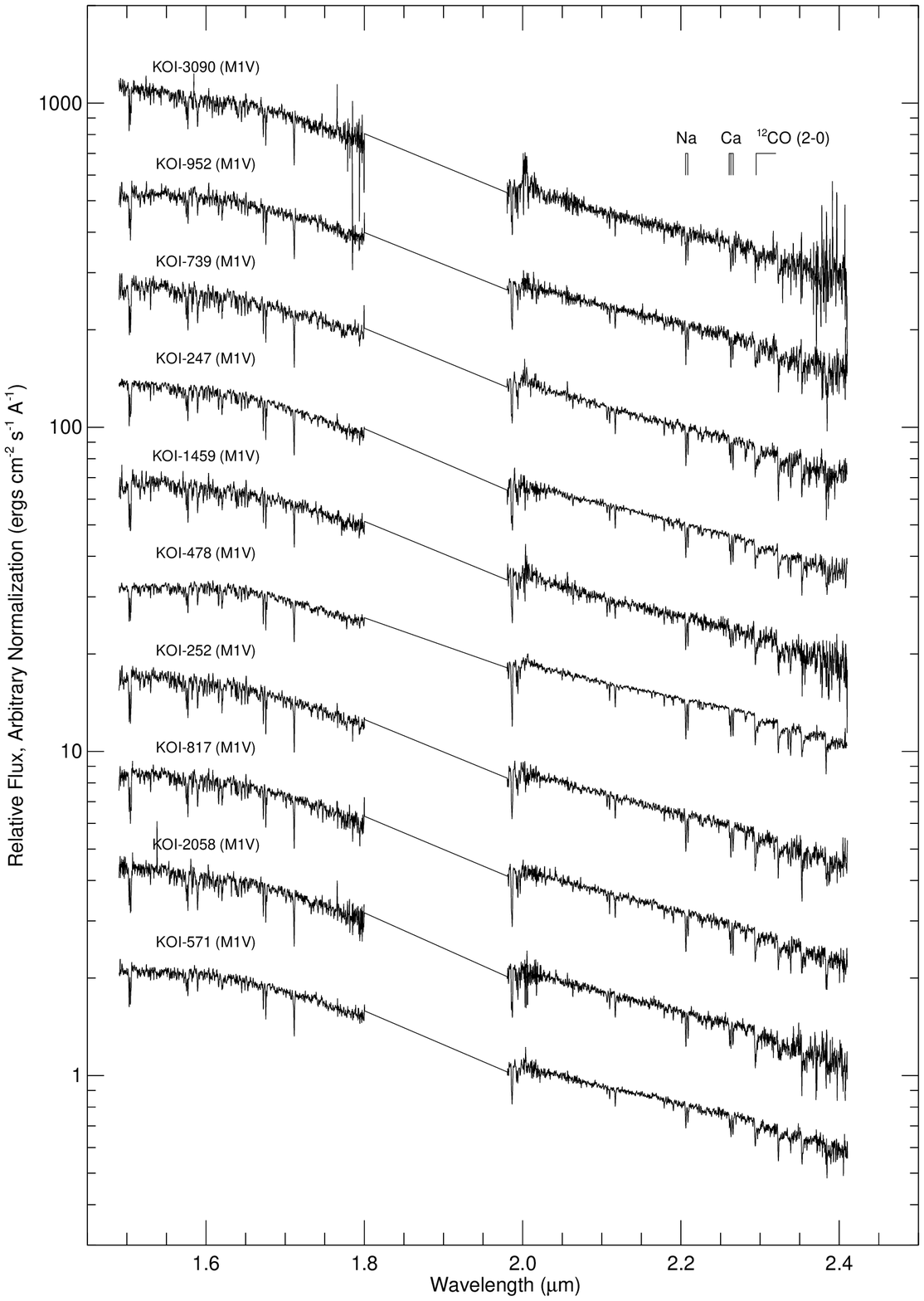}
\caption{continued.  \spectra}
\end{center}
\end{figure*}

\addtocounter{figure}{-1}
\begin{figure*}[]
\begin{center}
\includegraphics[width=6.5in]{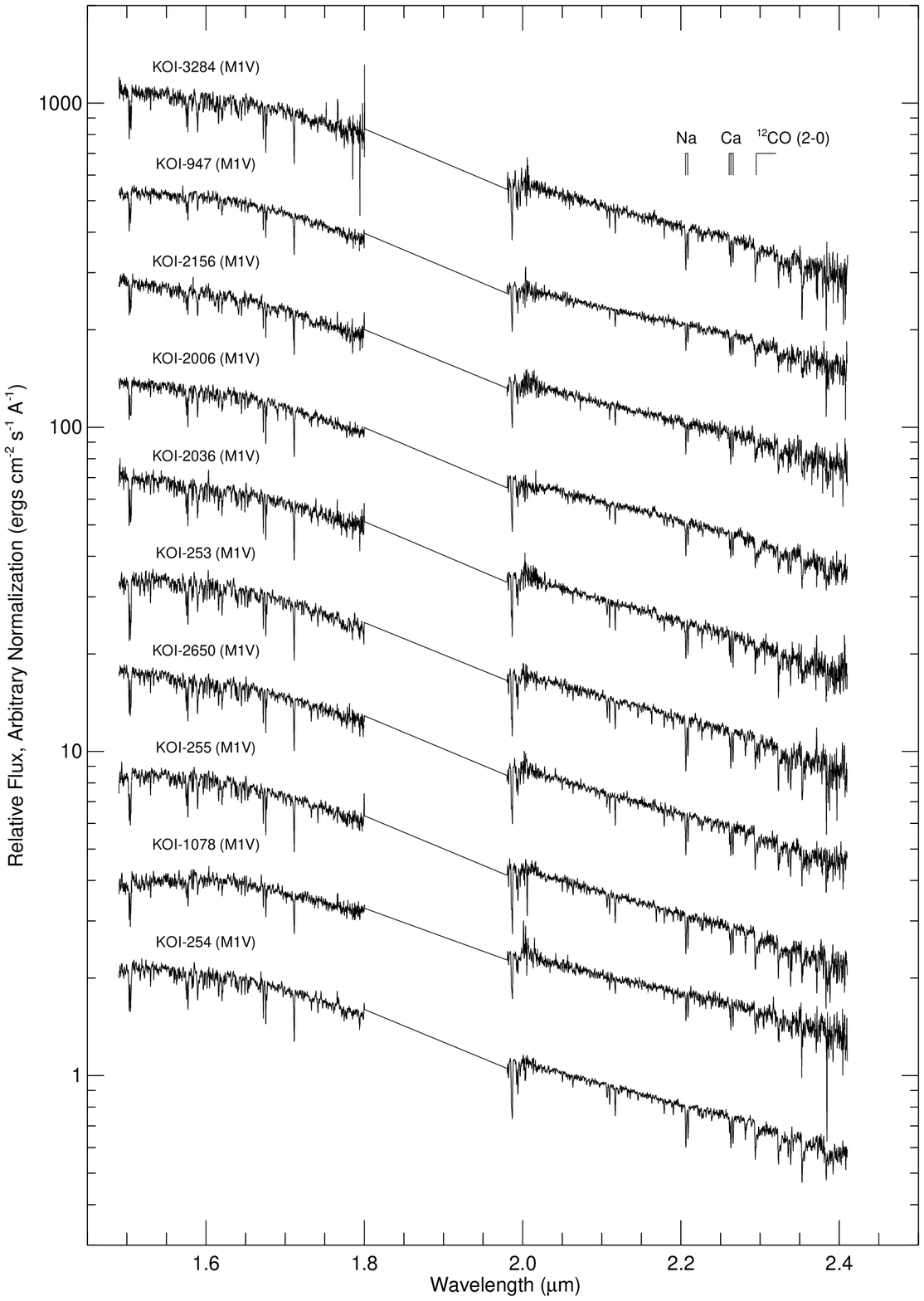}
\caption{continued.  \spectra}
\end{center}
\end{figure*}

\addtocounter{figure}{-1}
\begin{figure*}[]
\begin{center}
\includegraphics[width=6.5in]{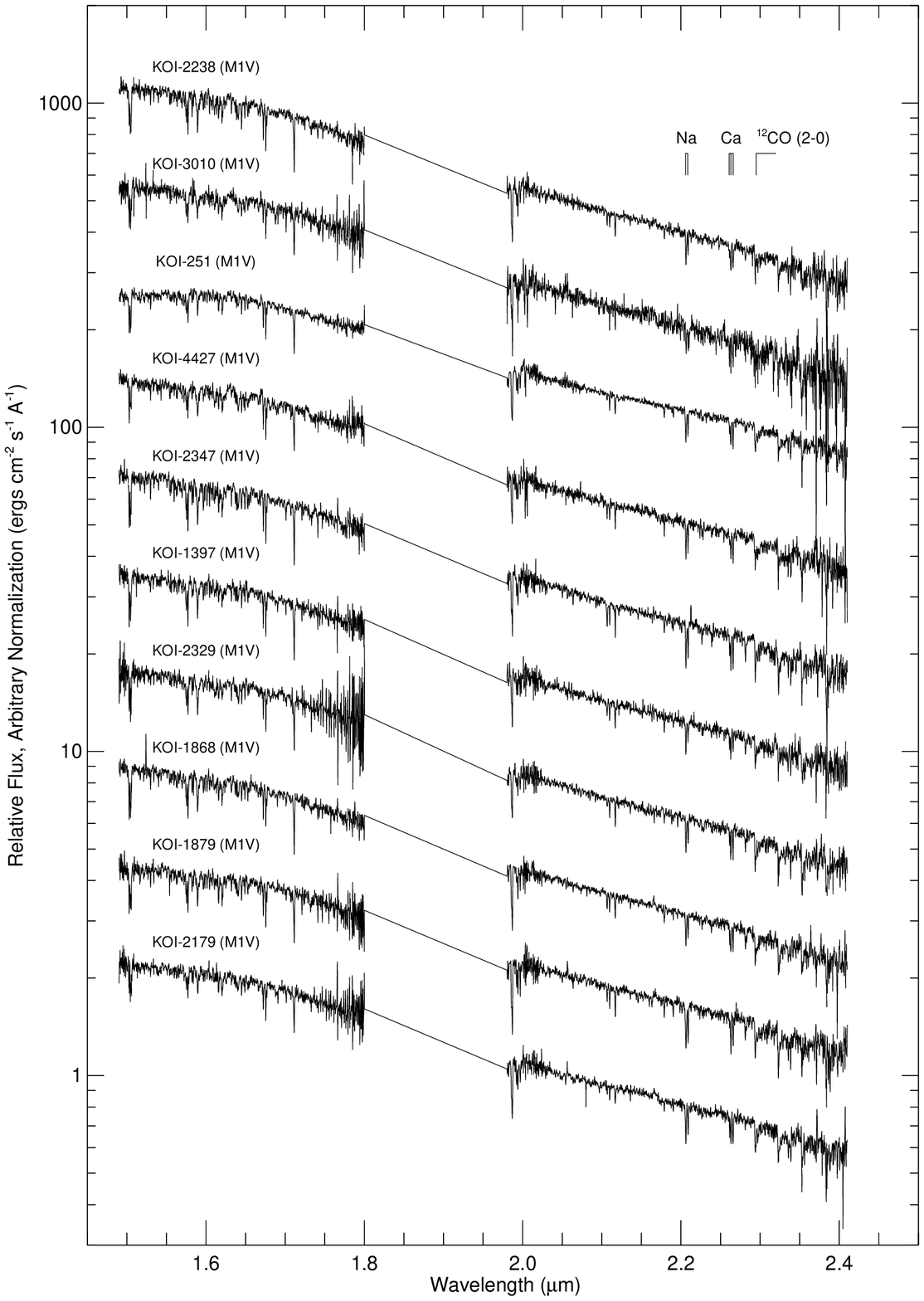}
\caption{continued.  \spectra}
\end{center}
\end{figure*}

\addtocounter{figure}{-1}
\begin{figure*}[]
\begin{center}
\includegraphics[width=6.5in]{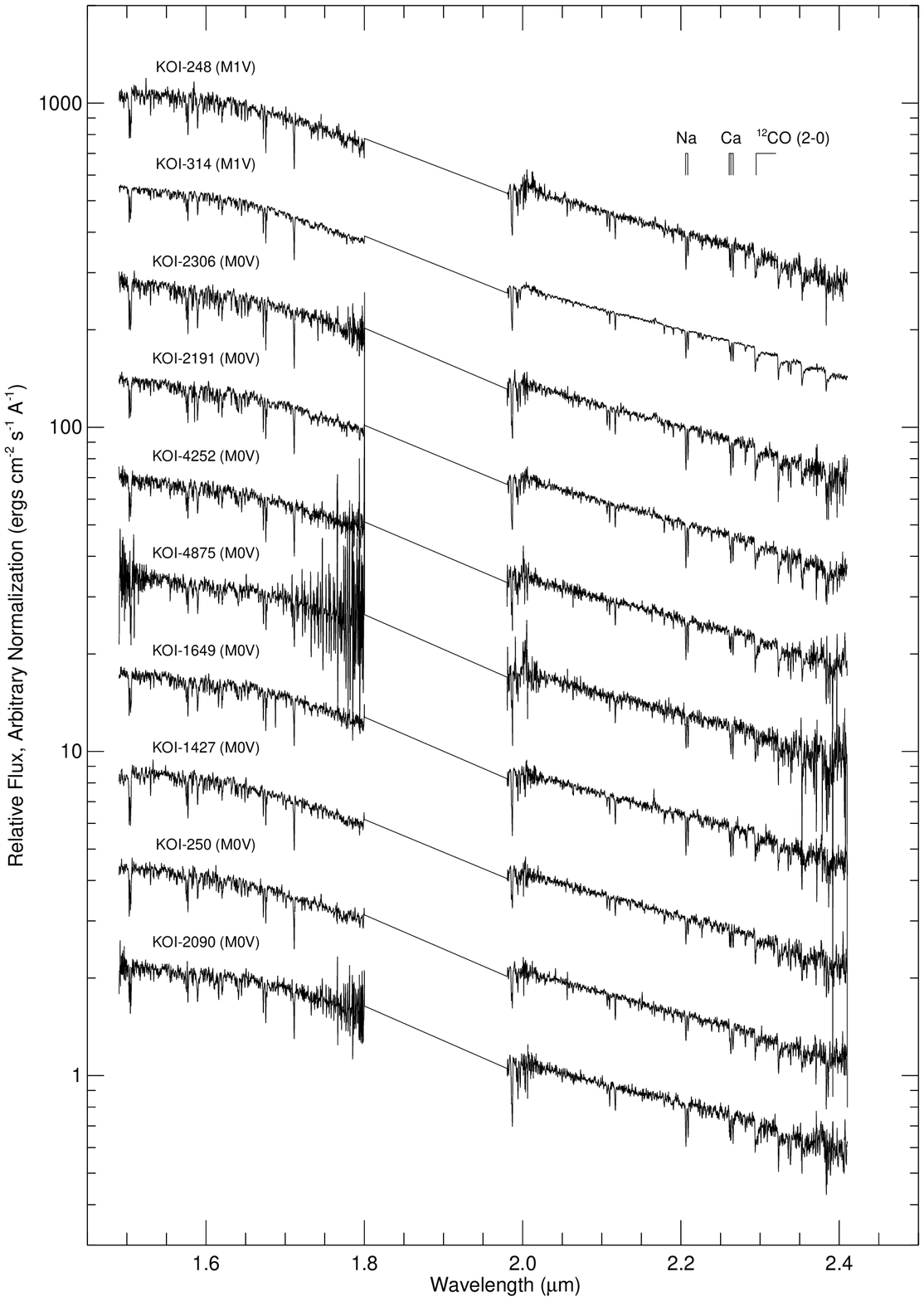}
\caption{continued.  \spectra}
\end{center}
\end{figure*}

\addtocounter{figure}{-1}
\begin{figure*}[]
\begin{center}
\includegraphics[width=6.5in]{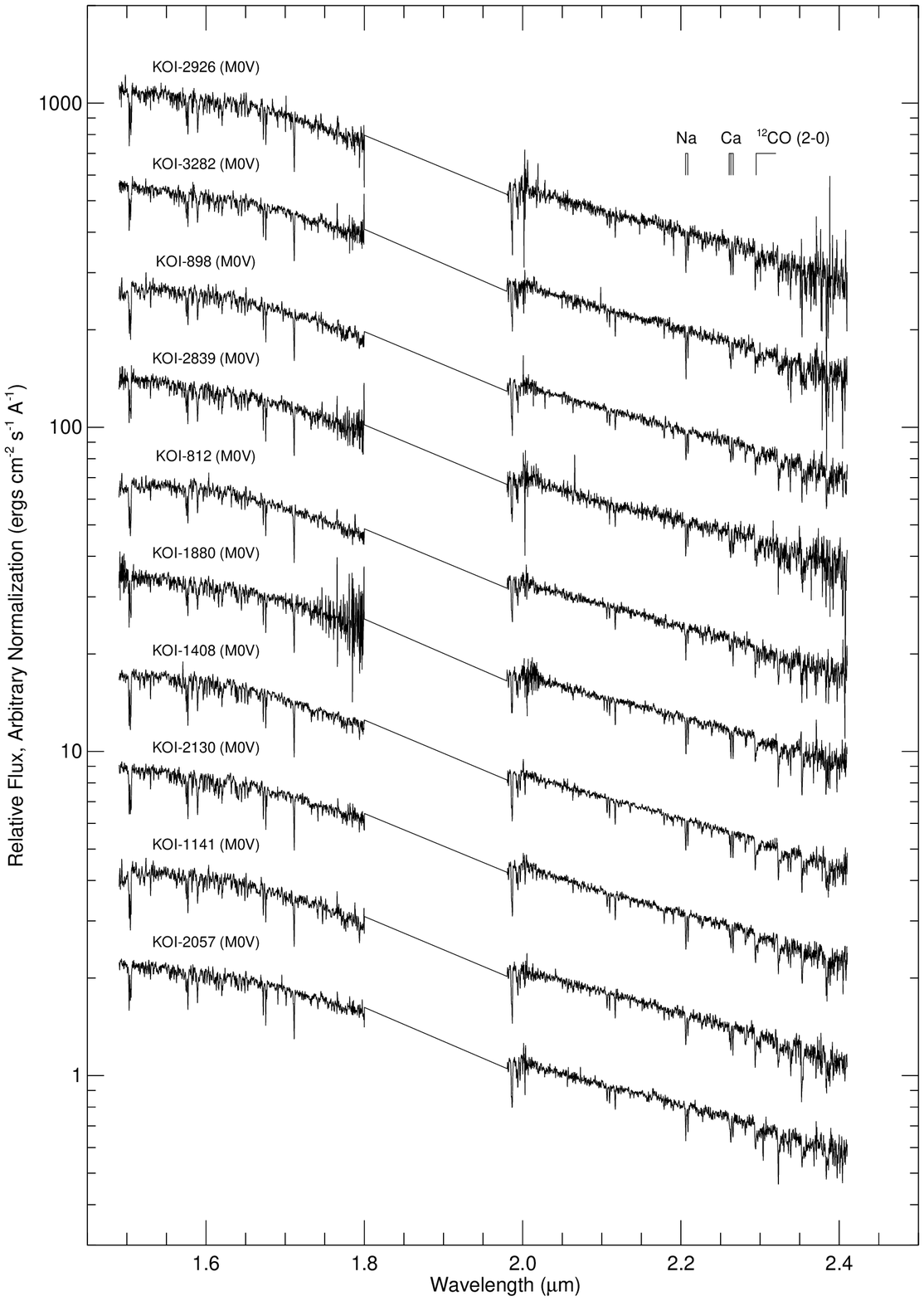}
\caption{continued.  \spectra}
\end{center}
\end{figure*}

\addtocounter{figure}{-1}
\begin{figure*}[]
\begin{center}
\includegraphics[width=6.5in]{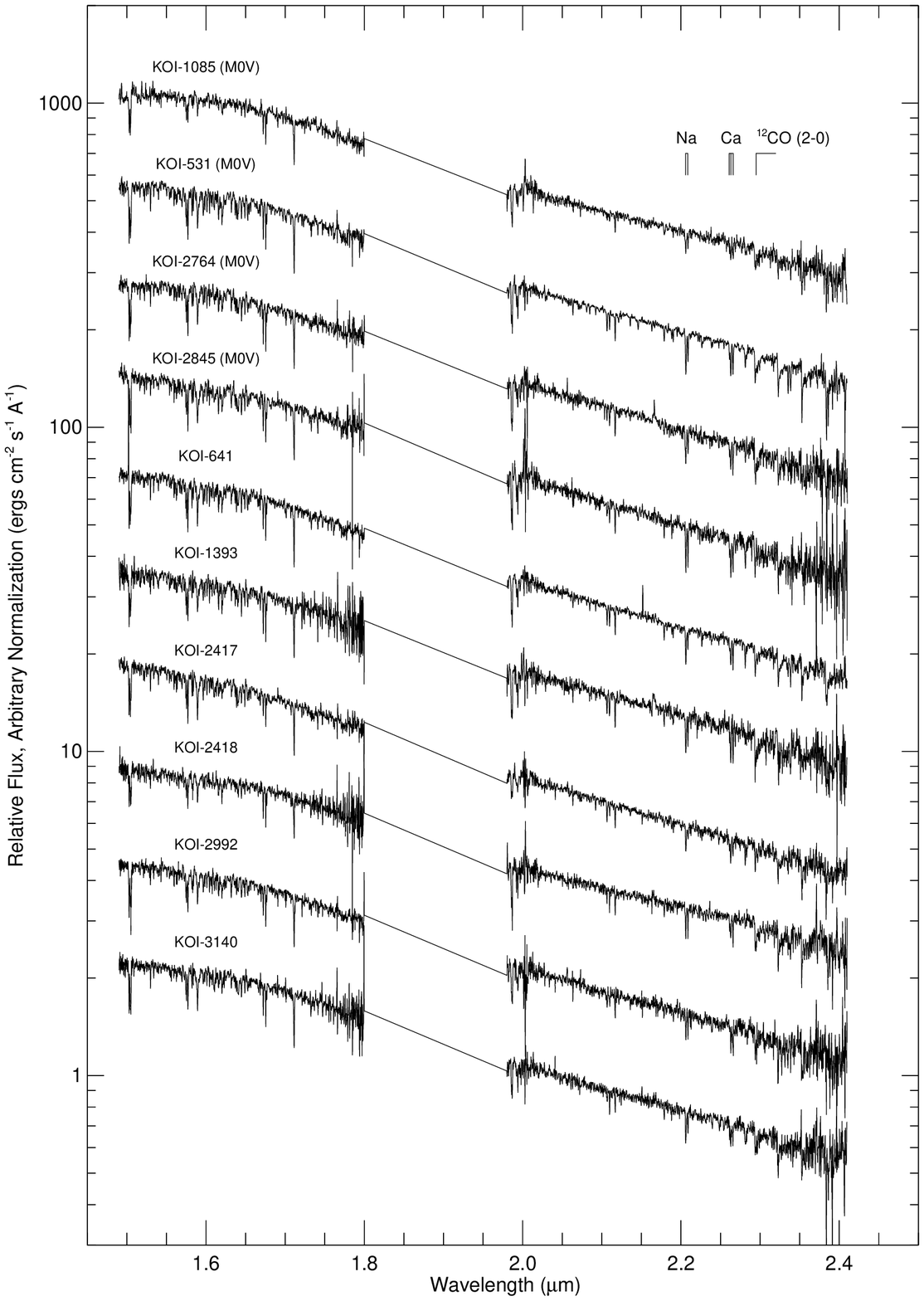}
\caption{continued.  \spectra}
\end{center}
\end{figure*}

\addtocounter{figure}{-1}
\begin{figure*}[]
\begin{center}
\includegraphics[width=6.5in]{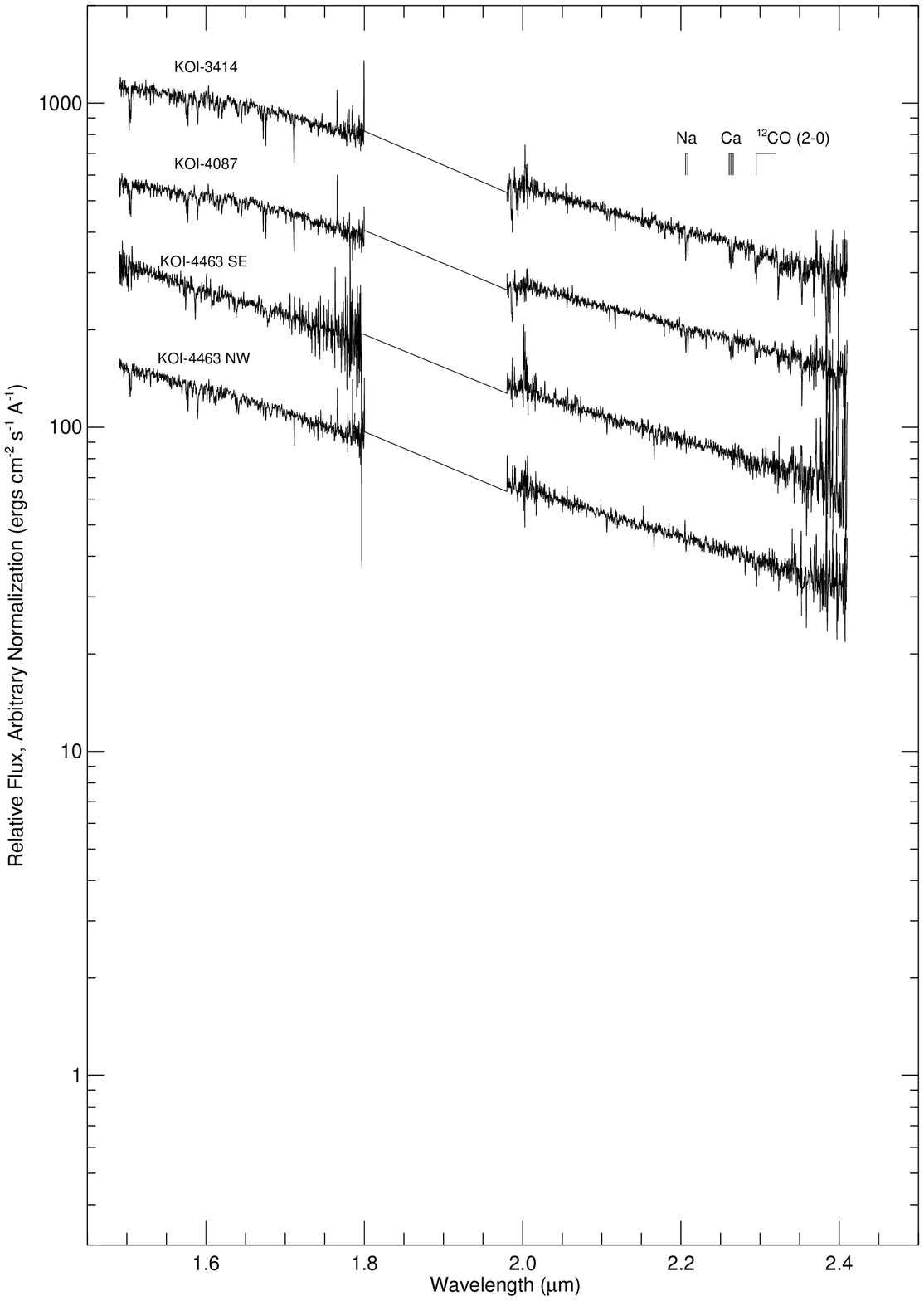}
\caption{continued.  \spectra}
\end{center}
\end{figure*}

\begin{figure*}[]
\begin{center}
\includegraphics[width=6.5in]{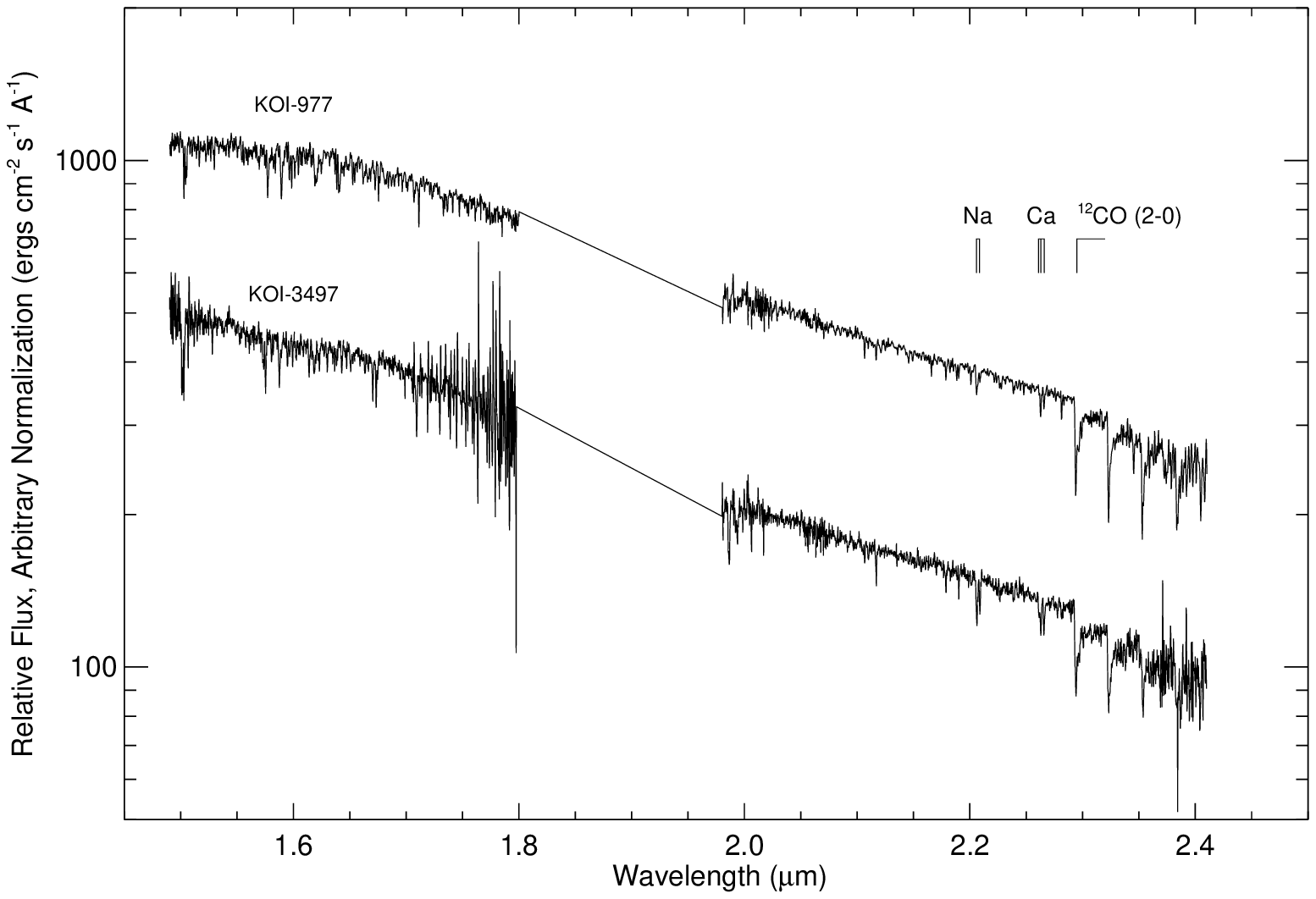}
\caption{$H$- and $K$ band spectra of KOI 977 and KOI 3497.  The spectrum of KOI 977 is consistent with a giant star: deep CO lines a 2.31 $\mu$m and shallow Na and Ca lines in $K$ band.  \citet{Mann2012} and \citet{Dressing2013} also conclude that KOI 977 is a giant star.  KOI 3497 is more interesting: it has deep 2.31 $\mu$m CO lines but also deep Ca and Na lines.  Adaptive optics imaging reveals that KOI 3497 is in fact two stars; however, their spectral types have not been confirmed.  The spectra are available for download online using the Data Behind the Figure feature.\label{giant_spectra}}
\end{center}
\end{figure*}

\acknowledgements

We would like to thank Andrew Mann and Ryan Terrien for helping us implement their respective methods for measuring metallicities with infrared spectra.  We would like to thank the staff at Palomar Observatory for providing support during our many observation runs, including Bruce Baker, Mike Doyle, Jamey Eriksen, Carolyn Heffner, John Henning, Steven Kunsman, Dan McKenna, Jean Mueller, Kajsa Peffer, Kevin Rykoski, and Greg van Idsinga. 

The Robo-AO system is supported by collaborating partner institutions, the California Institute of Technology and the Inter-University Centre for Astronomy and Astrophysics, and by the National Science Foundation under Grant Nos. AST-0906060 and AST-0960343, by the Mount Cuba Astronomical Foundation, by a gift from Samuel Oschin.

J.B. would like to thank Mr. and Mrs. Kenneth Adelman for providing funding for her 2012 Alain Porter Memorial Summer Undergraduate Research Fellowship.  A.V. is supported by the National Science Foundation Graduate Research Fellowship under Grant No. DGE1144152.  J.A.J. is supported by generous grants from the David and Lucile Packard Foundation and the Alfred P. Sloan Foundation.  This research has made use of the NASA Exoplanet Archive, which is operated by the California Institute of Technology, under contract with the National Aeronautics and Space Administration under the Exoplanet Exploration Program.  Some of the Palomar 200-inch Telescope time was provided by Cornell University.  P.S.M. acknowledges support for this work from the Hubble Fellowship Program, provided by NASA through Hubble Fellowship grant HST-HF-51326.01-A awarded by the STScI, which is operated by the AURA, Inc., for NASA, under contract NAS 5-26555.  C.B. acknowledges support from the Alfred P. Sloan Foundation

{\it Facilities:} \facility{Kepler}, \facility{PO:Hale (TripleSpec)}, \facility{PO:1.5m (Robo-AO)}

\clearpage
\bibliographystyle{apj}
\bibliography{Cool_koi_vi}

\end{document}